\documentclass[prx,amsmath,amssymb,notitlepage,nofootinbib,eqsecnum,longbibliography]{revtex4-2}

\pdfoutput=1

\usepackage{amsmath,amssymb,bm} 
\usepackage{graphicx}

\usepackage{color}
\definecolor{darkblue}{rgb}{0.,0.,0.4}
\definecolor{darkred}{rgb}{0.5,0.,0.}
\definecolor{BlueViolet}{RGB}{138,43,226}
\definecolor{SkyBlue}{RGB}{30,144,255}
\definecolor{DarkGreen}{RGB}{0,100,0}
\usepackage[pdftex,colorlinks=true,linkcolor=darkblue,citecolor=blue,urlcolor=blue]{hyperref}

\usepackage[all]{hypcap}
\usepackage[caption=false]{subfig}

\usepackage{amsfonts}
\usepackage{amsthm}
\usepackage{xcolor}

\newcommand{\bqa}{\begin{eqnarray}} 
\newcommand{\eqa}{\end{eqnarray}}
\newcommand{\nn}{\nonumber \\}

\newcommand{\ncmd}{\newcommand}

\ncmd{\lt}{\left}
\ncmd{\rt}{\right}
\ncmd{\tr}[1]{\mbox{Tr}\lt({#1}\rt)}
\ncmd{\half}{\frac{1}{2}}
\ncmd{\eps}{\epsilon}
\ncmd{\veps}{\epsilon}
\ncmd{\dgr}{\dagger}
\ncmd{\sig}{\sigma}
\ncmd{\gam}{\gamma}
\ncmd{\rtarw}{\rightarrow}
\ncmd{\Rt}{\Rightarrow}
\ncmd{\abs}[1]{\lt|{#1}\rt|}
\ncmd{\avg}[1]{\lt<{#1}\rt>}
\ncmd{\dl}{\delta}
\ncmd{\Dl}{\Delta}
\ncmd{\sgn}[1]{\mbox{sgn}\lt(#1\rt)}
\ncmd{\kap}{\kappa}
\ncmd{\wtil}[1]{\widetilde{#1}}
\ncmd{\thrfr}{\therefore}
\ncmd{\eq}[1]{Eq. \eqref{#1}}
\ncmd{\fig}[1]{Fig. \ref{#1}}
\ncmd{\Lam}{\Lambda}
\ncmd{\lam}{\lambda}
\ncmd{\dow}{\partial}
\ncmd{\ordr}[1]{\mathcal{O}\lt(#1\rt)}
\ncmd{\dsty}{\displaystyle}
\ncmd{\alert}[1]{\color{red}{#1}}
\ncmd{\mc}{\mathcal}
\ncmd{\mbf}[1]{\mathbf{#1}}
\ncmd{\derv}[2]{\frac{d{#1}}{d{#2}}}
\ncmd{\pderv}[2]{\frac{\partial{#1}}{\partial{#2}}}
\ncmd{\sub}[1]{{\,\dsty{\!_{#1}}}}
\ncmd{\step}[1]{\Theta\lt(#1\rt)}
\ncmd{\td}{\tilde} 
\ncmd{\what}{\widehat}
\ncmd{\om}{\omega}
\ncmd{\Om}{\Omega}
\ncmd{\vrho}{\varrho}
\ncmd{\vsig}{\varsigma}
\ncmd{\vkap}{\varkappa}
\ncmd{\nnum}{\nonumber}
\ncmd{\comment}[1]{{\color{red}{#1}}}
\definecolor{new_color}{RGB}{150,0,150}

\newcommand\redout{\bgroup\markoverwith
{\textcolor{red}{\rule[.5ex]{2pt}{0.8pt}}}\ULon}

\newcommand{\beq}{\begin{equation}}
\newcommand{\eeq}{\end{equation}}

\renewcommand{\vec}[1]{{\mathbf{#1}}}

\def\be{\begin{eqnarray}}
\def\ee{\end{eqnarray}}

\usepackage[papersize={8.5in,11in}]{geometry}
\begin{document}

\title{Scaling behaviour and superconducting instability in anisotropic non-Fermi liquids}

\author{Ipsita Mandal\\
\vspace{0.7cm}
{\normalsize{Perimeter Institute for Theoretical Physics,\\
31 Caroline St. N., Waterloo ON N2L 2Y5, Canada}}
}

\begin{abstract}
We study the scaling behaviour of the optical conductivity $(\sigma)$, free energy density $(F)$, and shear viscosity of the quantum critical point associated with the spin density wave phase transition for a two-dimensional metallic system with $C_2$ symmetry. A non-Fermi liquid behaviour emerges at two pairs of isolated points on the Fermi surface, due to the coupling of a bosonic order parameter to fermionic excitations at those so-called ``hot-spots''. We find that near the hot-spots, $\sigma$ and $F$ obey the scalings expected for such an anisotropic system, and the direction-dependent viscosity to entropy density ratio is not a universal number due to the anisotropy. Lastly, we also estimate the effect of the fermion-boson coupling at the hot-spots on superconducting instabilities.
\end{abstract}

\maketitle

\tableofcontents

\section{Introduction}

The ``strange metal'' phase observed in numerous correlated electron compounds, for example the cuprates, are unconventional metallic states that cannot be studied using the framework of the Landau Fermi liquid theory, as the quasiparticle excitations get destroyed due to their coupling with some gapless boson. There have been intensive efforts to study such ``non-Fermi liquid'' states \cite{holstein,reizer,leenag,HALPERIN,polchinski,ALTSHULER,YBKim,nayak,lawler1,lawler2,SSLee,metlsach1,metlsach,
chubukov1,Chubukov,chubukov3,mross,Jiang,Shouvik1,Lee-Dalid,shouvik2,patel1,sur16,ips1,Max,patel2,ips4,ips-sc,ips5,peter}. These states may involve the gapless bosons carrying either (1) zero momentum, such as the Ising-nematic critical point \cite{metlsach1,ogankivfr,metzner,delanna,delanna2,kee,lawler1,lawler2,rech,wolfle,maslov,quintanilla,yamase1,yamase2,halboth,
jakub,zacharias,YBKim,huh,Lee-Dalid,ips1,patel2,ips4,ips-sc,ips5} and nonrelativistic fermions coupled with an emergent gauge field \cite{MOTRUNICH,LEE_U1,PALEE,MotrunichFisher,ips2,ips3}; or (2) non-zero momenta, such as the spin density wave (SDW) and charge density wave (CDW) critical points \cite{metlsach,chubukov1,Chubukov,shouvik2,patel1,sur16,peter}.

Recently, a ``co-dimensional regularization scheme'' has been developed for a perturbatively controlled study of the SDW critical point in two-dimensional metals with four-fold ($C_4$) \cite{shouvik2} and ($C_2$) \cite{sur16} symmetries, by embedding the one-dimensional Fermi surface in a higher dimensional space. These are non-Fermi liquid systems where the critical theory is described by isolated points called ``hot-spots'', such that a bosonic order parameter is coupled to fermionic excitations at four (two) pairs of hot-spots around the Fermi surface with $C_4$ ($C_2$) symmetry. In the second case, the $C_4$-symmetric metallic state is explicitly or spontaneously broken to a $C_2$-symmetric one \cite{teifler,chu,chuang,fujita,ando,fink,hinkov,kivelson,Chubukov2015}, and
an anisotropic non-Fermi liquid emerges when the system undergoes a continuous density wave transition \cite{chu2,kasahara,zhou,lu,helm}. 

The hot-spot contribution to optical conductivity and finite temperature free energy density for the $C_4$-symmetric SDW critical point has been found in Ref.~\cite{patel1} using the regularization scheme of Ref.~\cite{shouvik2}, where the authors concluded that hyperscaling is obeyed near the hot-spots \footnote{In a recent work \cite{peter}, the authors have employed a non-perturbative treatment of the problem and found that there is hyperscaling violation for the free energy density in two spatial dimensions.}. This is expected for non-Fermi liquids arising from the interaction of the Fermi surface with bosons carrying non-zero momentum, where all the hot-spots exhibit an isotropy. In the present work, we compute the optical conductivity ($\sigma$) and free energy density for the anisotropic $C_2$-symmetric case, using the $\epsilon$-expansion of Ref.~\cite{sur16}. Furthermore, we calculate the shear viscosity ($\eta$) and find the scaling behaviour of the ratio between $\eta$ and entropy density ($s$). 
One can carry out a Boltzmann
analysis directly in $d = 2$, which will be very similar to the $C_4$-symmetric SDW case considered in Ref.~\cite{patel1}.
From such computations, it can be shown that the leading order temperature ($T$) dependence of the “quantum critical”
conductivity ($\sigma_Q$) has the same form as the frequency ($\omega$) dependence of $\sigma$. The $T$-dependence of the DC viscosity can also be inferred from the frequency dependent shear viscosity computed from the field theory. Lastly, we also estimate the effect of the fermion-boson coupling at the hot-spots on superconducting instabilities.

The paper is organized as follows:  In Sec.~\ref{model}, we review the co-dimensional regularization procedure devised in Ref.~\cite{sur16} to obtain a perturbative control of the $C_2$-symmetric SDW quantum critical point. In Sec.~\ref{optical}, we compute the scaling of the optical conductivity with frequency. Sec.~\ref{free-energy} deals with the calculation of the finite temperature free energy density. In Sec.~\ref{pairing}, we address the question whether the fermion-boson coupling results in an enhancement of the instability of four-fermion interactions to superconducting pairing. The expressions for the direction-dependent viscosity to entropy density ratios have been derived in Sec.~\ref{etabys}. We conclude with a summary and outlook in Sec.~\ref{summary}. The detailed computation of the current-current correlators has been shown in the appendix.


\section{Model}
\label{model}

\begin{figure}[]
\centering
\includegraphics[width = 0.25\columnwidth]{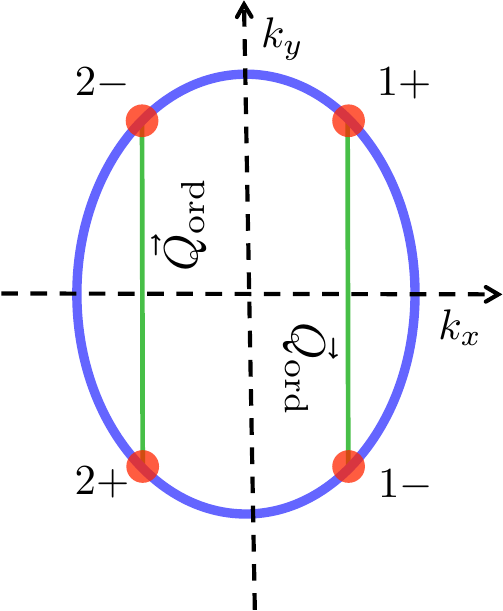}
\caption{\label{FS-scheme}
The anisotropic Fermi surface in two-dimensional momentum space spanned by $\vec k = (k_x, k_y)$. At the quantum critical point, SDW fluctuations induce strong scatterings 
between electrons near four hot-spots denoted by red dots. $\vec Q_{ord}$ represents the ordering vector which intersects these hot-spots.
}
\end{figure}

The action describing the fermions confined to two spatial dimensions and interacting with an SDW order parameter is given by \cite{sur16}:
\begin{align}
 S =& \sum_{j=1}^{N_f} \sum_{\sigma =1}^{N_c} \sum_{l=1}^{2} \sum_{m = \pm} \int \frac{d^3 k}{(2\pi)^3} ~ \psi_{l,m,j,s}^{*}(k)
\lt( ik_0 + \mc{E}_{l,m}(  k) \rt) \psi_{l,m,j,s}(k)  \nn
&
+ \frac{1}{4} \int \frac{d^3 q}{(2 \pi)^3} ~ 
\lt( q_0^2 + q_x^2 + c^2 q_y^2 \rt) \tr{\Phi(-q) \Phi(q)} \nn
&   + \frac{ g }{\sqrt{N_f}} \sum_{j=1}^{N_f} \sum_{l=1}^{2} \sum_{\sigma, \sigma' = 1}^{N_c} \int \frac{d^3 k}{(2 \pi)^3} 
\frac{d^3 q}{(2 \pi)^3}  
\lt[ \psi_{l,+,j,\sigma }^{*}(k+q)~ \Phi_{ \sigma,\sigma'}(q)  ~\psi_{l,-,j,s'}(k) + \mbox{h.c.} \rt] \nn
&   + \frac{1}{4} \int \frac{d^3 q_1}{(2\pi)^3} \frac{d^3 q_2}{(2\pi)^3} \frac{d^3 q_3}{(2\pi)^3}
~ \Bigl[ u_{1;0} ~ \tr{\Phi(-q_1 + q_2) \Phi(q_1)} ~ \tr{ \Phi(-q_3 - q_2) \Phi(q_3)} 
\nn &\qquad \qquad \qquad \qquad \qquad \, \quad
+ u_{2;0} ~ \tr{\Phi(-q_1 + q_2) \Phi(q_1) \Phi(-q_3 - q_2) \Phi(q_3)} \Bigr] ,
\label{dw-action-2d}
\end{align}
where the $ \psi_{l,m,j,\sigma}(k)$'s describe the electrons with momenta near the four hot-spots, 
labelled by $ (l=1,2, \,m=\pm )$, as shown in \fig{FS-scheme}. The labels
$j = 1,2,..,N_f$ and $\sigma=1,2,..,N_c$ represent the flavour and spin indices respectively, with
the $SU(2)$ spin generalized to $SU(N_c)$. 
The parameter $N_f$ is an extra flavour which can arise from degenerate bands with the $SU(N_f)$ symmetry. 
The $N_c \times N_c$ matrix field $\Phi(q)$ represents the SDW mode of frequency $q_0$ and momentum $\vec Q_{ord}+ \vec q$.
Furthermore, $\mc{E}_{1,+}(  k) = - \mc{E}_{2,+}(  k) = v \, k_x + k_y$,  $\mc{E}_{1,-}(  k) = - \mc{E}_{2,-}(  k) = v\, k_x - k_y$ and $( g ,  u_{1;0},  u_{2;0}  )$ are the coupling constants. The parameters $v \equiv \dfrac{v_x}{c_x}$ and $c \equiv \dfrac{c_y}{v_y}$ represent the relative velocities between electron and boson in the two directions.

In order to carry out dimensional regularization, the original $(2+1)$-dimensional theory is promoted to a $(d+1)$-dimensional theory which describes the one-dimensional Fermi surface embedded in a $d$-dimensional momentum space. 
We define new spinors $\Psi_{+,j,\sigma}(k) = \lt( \psi_{1,+,j,\sigma}(k), \, \psi_{2,+,j,\sigma}(k) \rt)^T$, $\Psi_{-,j,\sigma}(k) = \lt( \psi_{1,-,j,\sigma}(k), \, - \psi_{2,-,j,\sigma}(k) \rt)^T$, and $\bar \Psi_{n,j,\sigma} = \Psi_{n,j,\sigma}^{\dag} \gam_0$.
Adding $(k_1, \ldots, k_{d-2} )$ as the extra $(d-2)$ dimensions perpendicular to the Fermi surface (co-dimensions), the new action is given by:
\begin{align}
 S = &\sum_{j=1}^{N_f} \sum_{\sigma =1}^{N_c} \sum_{n = \pm} \int dk ~ \bar \Psi_{n,j,\sigma}(k) \lt( i\, \mbf{K} \cdot \mbf{\Gamma} + i \,\varepsilon_{n}(k ) ~ \gamma_{d-1} \rt) \Psi_{n,j,\sigma}(k) \nn
 &
 + \frac{1}{4} \int dq ~ \lt( |\mbf{Q}|^2 + q_x^2 + c^2 q_y^2 \rt) ~ \tr{\Phi(-q) \, \Phi(q)} \nn
&  + i ~ \frac{g\, \mu^{\frac{3-d}{2}} }{\sqrt{N_f}} \sum_{j=1}^{N_f}  \sum_{\sigma,\sigma' = 1}^{N_c} \int dk ~ dq ~  
\lt[ \bar \Psi_{+,j,\sigma}(k+q)~ \gam_{d-1} ~ \Phi_{\sigma,\sigma'}(q) ~\Psi_{-,j,\sigma'}(k) - \mbox{h.c.} \rt] \nn
&   + \frac{ \mu^{3-d} }{4} \int dq_1 \,  dq_2 \,  dq_3 ~ \Bigl[ u_{1;0} ~ \tr{\Phi(-q_1 + q_2) \Phi(q_1)} ~ \tr{ \Phi(-q_3 - q_2) \Phi(q_3)} \nn
&
\qquad \qquad \qquad   \qquad \,
+ u_{2;0} ~ \tr{\Phi(-q_1 + q_2) \Phi(q_1) \Phi(-q_3 - q_2) \Phi(q_3)} \Bigr],
\label{gen-d}
\end{align}
where $k \equiv (\mbf{K}, \vec k)$, $dk \equiv \dfrac{d^{d+1} k}{(2\pi)^{d+1}}$ and $\mbf{K} \equiv (k_0, k_1, \ldots, k_{d-2})$.
We have introduced a mass scale $\mu$ to make the coupling constants $g$, $u_{1;0}$ and $u_{2;0}$ dimensionless. The vector $\mbf{\Gamma} \equiv (\gam_0, \gam_1, \ldots, \gam_{d-2})$ has the first $(d-1)$ $\gamma$-matrices as its components. Since in real systems, $d $ lies between $2$ and $3$, 
we will consider only the $2 \times 2$ gamma matrices so that the corresponding spinors always have two components. We will use the representation
where
 \beq
 \Gamma_0  \equiv \gamma_0= \sigma_y \,, \quad \gamma_{d-1} = \sigma_x \,
 \eeq
are fixed in general dimensions.
The dispersions of the spinors, $\varepsilon_{n }(  k) = v \, k_x + n\, k_y$, are inherited from the two dimensional dispersions $( \mc{E}_{1,\pm}(  k), \, \mc{E}_{2 , \pm }( k))$. 
The original $(2+1)$-dimensional action in \eq{dw-action-2d} can be obtained by setting $d=2$.

The fermionic and bosonic Green's functions are given by:
\begin{align}
G_n(k)&=-i \,\frac{\mathbf{\Gamma}\cdot\mathbf{K}+\gamma_{d-1} \, \varepsilon_n( k )}{\vert \mathbf{K}\vert ^2+\varepsilon_n^2( k )} \, ,\quad
D(q) =\frac{1}{\vert\mathbf{Q}\vert^2+q_x^2+c^2 \, q_y^2} \, ,
\end{align}
respectively. 
The fixed point of the model is characterized by:
\begin{align}
v=v_* = \frac{ N_c\, N_f}{2\, (N_c^2-1)}\,, \quad
g^2= g^2_*=\frac{8\, \pi N_c\, N_f}{(N_c^2-1)}\,(z_\tau-1) \,, \quad
z_\tau=1+\frac{ 2 \left ( N_c^2 -1 \right ) + N_c \, N_f   }
{2 \, \big \lbrace  
2 \left ( N_c^2 -3 \right ) + N_c \, N_f
\big \rbrace }
\, \epsilon\,, \quad
\quad 
\end{align}
at one-loop order, in an expansion in the parameter $\epsilon = 3-d$, where $z_\tau$ is the dynamical critical exponent.


\section{Renormalized optical conductivity}
\label{optical}

The current densities in the $x$ and $y$ directions are given by:
\be
\label{eqn:jexprs}
J_x &=& i \, v  \sum_{j=1}^{N_f}  \sum_{\sigma=1}^{N_c} \sum_{ n=\pm}
\bar \Psi_{n,j,\sigma } \gamma_{d-1} \Psi_{n,j,\sigma } \,,\quad
J_y = i    \sum_{j=1}^{N_f}  \sum_{\sigma=1}^{N_c} \sum_{ n=\pm} n \, \bar \Psi_{n,j,\sigma } \gamma_{d-1} \Psi_{n,j,\sigma } \,.
\ee
In order to obtain the optical conductivity, we need to calculate the expectation values $\langle J_x J_x \rangle$ and $\langle J_y J_y \rangle$.

\subsection{One-loop contributions to $\langle J_x J_x \rangle  $ and $\langle J_y J_y \rangle  $}

\begin{figure}
\centering
\includegraphics[width=0.32 \textwidth]{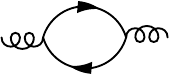} 
\caption{The one-loop diagram contributing to $\langle J_\mu J_\mu \rangle (\omega) $.}
\label{one-loop-fig}
\end{figure}

The current-current correlation functions at one-loop level are given by (Fig.~\ref{one-loop-fig}):
\be
\langle J_x J_x \rangle_{\text{1-loop}} (\omega) &=& v^2  \sum_{j=1}^{N_f}  \sum_{\sigma=1}^{N_c} \sum_{ n=\pm} \int dk \,
 \mathrm{Tr} \Big [ \gamma_{d-1} \, G_n (k+q) \,\gamma_{d-1} G_{n}( q) \Big] 
=   \frac{   v\,  N_f N_c \,  \omega^{1-\epsilon } }  { 32\, \pi    }
  \int d \varepsilon_{ - } (k) \,, \nn
\langle J_y J_y \rangle_{\text{1-loop}} (\omega) &=&     \sum_{j=1}^{N_f}  \sum_{\sigma=1}^{N_c} \sum_{ n=\pm} \int dk\,\mathrm{Tr} \Big [ \gamma_{d-1} \, G_n (k+q) \,\gamma_{d-1} G_{n}( q) \Big] 
=  \frac{    N_f N_c \,  \omega^{1-\epsilon } }  { 32 \, \pi \, v   }
  \int d \varepsilon_{ - }   (k) \,,\nn
\ee
where $ q = \left( \omega, 0, \ldots,0\right )$ and $d= 3-\epsilon $.
The calculational details have been shown in Appendix~\ref{one-loop-jj}.

\subsection{Two-loop contributions to $\langle J_x J_x \rangle  $ and $\langle J_y J_y \rangle  $}

\begin{figure}
        \centering
        \subfloat[][]
                {\includegraphics[width=0.32 \textwidth]{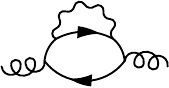} 
              { \label{fig:4vv1}}}
          \subfloat[][]
                {\includegraphics[width=0.32 \textwidth]{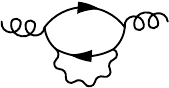}
		{\label{fig:4vv2}}}
		\subfloat[][]
                {\includegraphics[width= 0.34 \textwidth]{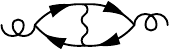}
		{\label{fig:4vv3}}}
        \caption{\label{loop2}The two-loop diagrams contributing to $\langle J_\alpha J_\alpha \rangle (\omega) $.}
\end{figure}
At two-loop order, the diagrams contributing to $\langle J_\alpha J_\alpha \rangle (\omega) $ are shown in Fig.~\ref{loop2}.
The first two diagrams have identical contributions and involve the fermion-self energy correction to $\langle J_\mu J_\mu \rangle (\omega) $. The last one corresponds to the vertex corrections. For $\alpha=x$, these can be written as:
\begin{align}
\label{jjSE}
\langle J_x J_x\rangle_{\text{SE}}(q)&=2   \,  v^2\sum_{j=1}^{N_f}\sum_{\sigma=1}^{N_c}\sum_{n=\pm}\int dk\,\mathrm{Tr}\left[  \,   \gamma_{d-1}\, G_n(k)\, \Sigma_{1,  n} (k)\, G_n(k)\,
  \gamma_{d-1 }  \, G_n(k+ q) \, \right],
\\
\langle J_x J_x\rangle_{\text{VC}}(q)&= i\, v^2  \sum_{j=1}^{N_f}\sum_{\sigma=1}^{N_c}\sum_{n=\pm}
  \int dk\,\mathrm{Tr}\left[  \,  \gamma_{d-1}\, G_n(k )\, \Xi_{  n }(k)\, G_n(k+ q) \, \right],\label{jjvert}
\end{align}
respectively.
Here, $ \Sigma_{1,  n} (k)$ and $ \Xi_{ n }(k)$ are the one-loop fermion self-energy and fermion-boson vertex corrections respectively, using the dispersion $\varepsilon_{  n }(  k)$.

As explained in details in Appendix~\ref{2loop-self}, the leading order term in $c$ is obtained only from $\langle J_xJ_x\rangle_{\text{SE}} (\omega)$. This is because the leading order term in $\langle J_x J_x\rangle_{\text{VC}}(\omega)$ is proportional to $c $.

The final result for $\langle J_xJ_x\rangle_{\text{SE}}$ is given by:
\begin{align}
\langle J_x J_x \rangle_{\text{SE}}(\omega)=
\begin{cases}
\frac{ -v\,  ( N_c^2 - 1)\,
\, g^2 \, \omega^{1 +\frac{1} {z_\tau }-\epsilon } }
{    128\, \pi^2 \,   \epsilon  }
 \left (
\frac{\mu}
{\omega}
\right )^{\epsilon}  & \mbox{for regions close to hot-spots} \,, \\
\frac{- v\,  ( N_c^2  - 1 )\,
\, g^2 \,\lambda \,  \omega^{1  -\epsilon } }
{  128\, \pi^2 \,   \epsilon  }
 \left (
\frac{\mu}
{\lambda}
\right )^{\epsilon} 
& \mbox{for regions far from hot-spots} \,,
\end{cases} 
\end{align}
to leading order in $\epsilon $ and $c$, where
$\lambda$ is a scale independent of $\omega$ and is of the order of $k_F \gg \omega $.
It is easy to see that $\langle J_yJ_y\rangle_{\text{SE}}(\omega) = \frac{\langle J_xJ_x\rangle_{\text{SE}}(\omega)} {v^2 } $.


\subsection{Scaling of optical conductivity}

Near the hot-spots, using $g^2_*=\frac{8\, \pi N_c\, N_f}{(N_c^2-1)}\,(z_\tau-1)$, the total contribution for $\langle J_xJ_x\rangle$ takes the form :
\begin{align}
& \langle J_xJ_x\rangle (\omega )
=
\langle J_x J_x \rangle_{\text{1-loop}} (\omega)
+ \langle J_xJ_x\rangle_{\text{SE}} (\omega)
+ \langle J_xJ_x\rangle_{VC}(\omega) + \langle J_xJ_x\rangle_{\text{counterterms}}(\omega) 
\nn
&
= \frac{   v_*\,  N_f N_c \,  \omega^{ 2-\epsilon } }  { 16 \, \pi    }
- 
\frac{   v\,  ( N_c^2 -  1)
\, g_*^2 \, \omega^{1 +\frac{1} {z_\tau }-\epsilon } }
{  128 \, \pi^2  }
\ln \left (
\frac{\mu}
{\omega}
\right ),  \nn
&=\frac{   v_*\,  N_f N_c \,  \omega^{ 2-\epsilon } }  { 16\, \pi    }
+ \frac{  v\,   N_f N_c \,
 (  z_\tau -1  ) \,  \omega^{1+\frac{1} {z_\tau }-\epsilon }}
{  16 \,  \pi   }
\ln \left (
\frac{\omega} {\mu} \right )\,,
\end{align}
where we have used the fact that the $1/\epsilon $ piece from $\langle J_xJ_x\rangle_{\text{SE}} (\omega)$ is cancelled by the corresponding counterterm.

We note that near the hot-spots, the momenta scale as $ k_y \sim \omega ^{\frac{ 1} { z_\tau} }$ and $ k_x \sim \omega ^{\frac{z_x} { z_\tau} } \sim v\, k_x $. This tells us that $v \sim v_* \, \omega ^{\frac{ 1- z_x } { z_\tau} }$ due to this anisotropic scaling. Using this information, we can rewrite  $\langle J_xJ_x\rangle (\omega )$ as
\begin{align}
& \langle J_xJ_x\rangle (\omega )=\frac{   v_* \,  N_f N_c \,  \omega^{ 2-\epsilon } }  { 16\, \pi    }
\Big [ 1+
 (  z_\tau -1  ) \,  \omega^{ \frac{1} {z_\tau }+\frac{ 1- z_x } { z_\tau}  -1  }
\ln \left (
\frac{\omega} {\mu} \right )
\Big ]  \nn
&
\simeq \frac{   v_* \,  N_f N_c \,  \omega^{ 2-\epsilon } }  { 16\, \pi    }
\Big [ 1+
  \,  \omega^{ \frac{1} {z_\tau }+\frac{ 1- z_x } { z_\tau}  -1 +(  z_\tau -1  ) }
\Big ] \,.
\end{align}
Using the fact that $(1-z_\tau)$ and $(1-z_x)$ are at least $\mathcal{O} (\epsilon )$, we argue that $\big [\frac{1-z_\tau} {z_\tau }+\frac{ 1- z_x } { z_\tau}   +(  z_\tau -1  ) \big ] \sim   \frac{ 1- z_x } { z_\tau}    $. Hence, the corresponding conductivity is given by
\begin{align}
& \sigma_{xx} (\omega )
\simeq  - \frac{   v_* \,  N_f N_c \,  \omega^{ 1-\epsilon } }  { 16\, \pi    }
\Big [ 1+
  \,  \omega^{ \frac{ 1- z_x } { z_\tau}  }
\Big ] \,,
\end{align}
setting $\mu=1 $.
Therefore the hot-spot contributions to the optical conductivity
scales with frequency as $ \sigma_{xx} (\omega ) \sim  \omega^{ 1-\epsilon + \frac{ 1- z_x } { z_\tau} } $, as expected from scaling arguments. Since $\langle J_yJ_y\rangle = \frac{\langle J_xJ_x\rangle} {v^2 } $, it immediately follows that $ \sigma_{y y } (\omega ) \sim  \omega^{ 1-\epsilon - \frac{ 1- z_x } { z_\tau} } $.

On the other hand, for points away from the hot-spots, the integral over $\varepsilon_-(k) $ has $\pm k_F $ as the bounds. In this case, we have:
\begin{align}
& \langle J_xJ_x\rangle (\omega )
= \frac{   v\,  N_f N_c \, k_F \,  \omega^{ 1-\epsilon } }  { 16 \, \pi    }
-
\frac{    v\,  ( N_c^2 - 1)
\, g^2 \, k_F \, \omega^{1  -\epsilon } }
{  128\,\pi^2 \,   \epsilon  }
\ln \left (
\frac{\mu}
{\lambda }
\right ),  \nn
&=\frac{   v\,  N_f N_c \, k_F \,  \omega^{ 1-\epsilon } }  { 16\, \pi    }
+ \frac{  v\,   N_f N_c \,
 (  z_\tau -1  ) \, k_F \, \omega^{1 -\epsilon }}
{  16 \,  \pi   }
\ln \left (
\frac{ k_F } {\mu} \right )\,,
\end{align}
implying that
\begin{align}
\sigma_{xx}  (\omega ) \sim \sigma_{yy}  (\omega ) \sim k_F \, \omega^{  - \epsilon} .
\end{align}

\section{Free energy at finite temperature}
\label{free-energy}

\begin{figure}[h!]
        \centering
\includegraphics[width=0.2 \textwidth]{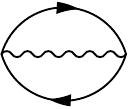}
\label{fig:freediag}
        \caption{Contribution to the interacting part of the free energy.}
\end{figure}

In this section, we compute the free energy density at a finite temperature $T > 0$ to leading order in $c$ and $\epsilon$. The free energy density receives contributions from three parts, namely, the free fermions, the free bosons, and the corrections due to the interactions between the two.

The contribution from the free fermionic part is given by:
\begin{align}
F_f^0(T) &=
2 \, N_f N_c 
\int\frac{ dk_x \, dk_y\,d^{1-\epsilon}\mathbf{\bar{K}}}
{(2 \, \pi)^{ 3-\epsilon}}
\, \Big [
T \sum_{p=\pm}
\ln \Big \lbrace \left(1+
e^{ p \sqrt{ \varepsilon_+^2(k) +\mathbf{\bar{K}}^2}/T}\right)
\Big \rbrace 
-  \sqrt{  \varepsilon_+^2(k) +\mathbf{\bar{K}}^2}  \, \Big ]\nn
&=\frac{  N_f N_c   } {v_* }
 \int\frac{ d \varepsilon_+(k) \, d  \varepsilon_-(k)\,d^{1-\epsilon}\mathbf{\bar{K}}}{(2 \, \pi)^{ 3-\epsilon}}
\, \Big [
T \sum_{p=\pm}
\ln \Big \lbrace \left(1+
e^{ p \sqrt{ \varepsilon_+^2(k) +\mathbf{\bar{K}}^2}/T}\right)
\Big \rbrace 
-  \sqrt{  \varepsilon_+^2(k)   +\mathbf{\bar{K}}^2} \,  \Big ],
\end{align} 
where we have subtracted the infinite contribution from the temperature-independent ground state energy.
Defining $\tilde k \, T =\sqrt{ \varepsilon_+^2(k) +\mathbf{\bar{K}}^2 } $, we thus have:
\begin{align}
F_f^0(T)  &=
\frac{ 2 \,\pi^{\frac{  2-\epsilon}{2}} \, N_fN_c \, T^{ 3 -\epsilon }  }
{(2\pi)^{3-\epsilon} \, v _*\, \Gamma\left(  \frac{ 2-\epsilon}{2} \right)}
\int d  \varepsilon_-(k)  \, d\tilde k\, \tilde k^{1-\epsilon} \left [
 \ln
\Big  \lbrace \left(1+e^{\tilde k}\right)\left(1+e^{-\tilde k}\right)  \Big \rbrace-\tilde k
\right ]\nn
&=
\frac{ 4 \,\pi^{\frac{ 2-\epsilon}{2}} \, N_fN_c \, T^{ 3 -\epsilon }  }
{(2\pi)^{3-\epsilon} \, v_* \, \Gamma\left(  \frac{ 2-\epsilon}{2} \right)}
\int  d  \varepsilon_-(k)  \, d\tilde k\, \tilde k^{1-\epsilon} \,
 \ln   \left(1+e^{ - \tilde k}\right) \nn
 &
 =\frac{ N_f N_c\, T^{3-\epsilon} \,\eta(3)}
{ 2\, \pi^2 \, v_* } \int  d  \varepsilon_-(k)
 =\frac{ 3\,N_f N_c\, T^{3-\epsilon} \,\zeta(3)}
{8\, \pi^2 \, v_*}  \int  d  \varepsilon_-(k) \,,
\end{align}
where $ \eta (u) = \frac{1}{\Gamma( u-1)} \int_0^\infty t^{u-2} \, \ln \left( 1+e^{-t}\right)$ is the Dirichlet eta function such that $ \eta (u)= \left( 1- 2^{1-u }\right) \, \zeta( u ) $.

For the free bosonic contribution, we have:
\begin{align}
F_b^0(T)  &=(1-N_c^2) \,T
\int\frac{ dq_x\,dq_y \, d^{1-\epsilon}\mathbf{\bar{Q}}}{(2 \, \pi)^{3-\epsilon}}
\ln \left(1-e^{-\sqrt{q_x^2+c^2q_y^2+\mathbf{\bar{Q}}^2}/T}\right).
\end{align}
Scaling \(q_y\to q_y/c\) and defining $\tilde  q \, T=\sqrt{q_x^2+q_y^2+\mathbf{\bar{Q}}^2  }$, we obtain:
\begin{align}
F_b^0(T)
&=\frac{ 2 \,\pi^{\frac{3-\epsilon}{2}} \,(1-N_c^2) \,T^{4-\epsilon}}
{(2\, \pi)^{3-\epsilon} \, c \,  \Gamma\left(  \frac{3-\epsilon}{2} \right)}
\int d\tilde q  \,\tilde  q^{2-\epsilon} \, \ln \left(1-e^{-\tilde q }\right)\nn
&=\frac{ 2 \,\pi^{\frac{3-\epsilon}{2}} \,( N_c^2 - 1) \,T^{4-\epsilon}\, \zeta(4)}
{(2\, \pi)^{3-\epsilon} \, c \,  \Gamma\left(  \frac{3-\epsilon}{2} \right)} 
=\frac{\pi^2  \, (N_c^2-1) \,T^{4-\epsilon}}
{ 90 \, c} \,,
\end{align}
where we have used the relation $\zeta (u) = -\frac{1}{\Gamma( u-1)} \int_0^\infty t^{u-2} \, \ln \left( 1-e^{-t}\right)$ for the Riemann zeta function.

The first order interaction correction to the free energy, as shown in Fig.~\ref{fig:freediag}, is given by:
\begin{equation}
\begin{split}
&F_{fb}(T)  \nn
	 &=  {g^2 \, (N_c^2 - 1)\, \mu^\epsilon \, T ^2}
 \sum_{\Omega_p, \,\omega_{p' } } \int \frac{ dq_x \, dq_y\,  d^{ 1-\epsilon} \vec{ \bar Q} \,dk_x \, dk_y\,  d^{ 1-\epsilon} \vec{ \bar K}  }   {  (2 \, \pi)^{6- 2\, \epsilon } }
 \frac{ 1 }
 { \Omega_p^2 +  \vec {\bar Q}^ 2  + q_x^2+c^2 \, q_y^2 } \nn
	&\qquad \qquad \qquad \qquad \qquad \qquad  \times \, 
	\sum_{n=\pm} \mathrm{Tr}\Big  [\gamma_{d-1} \, G_{n} ( \omega_{p' } + \Omega_p, 
\vec {\bar  K} + \vec {\bar  Q}, \vec k + \vec  q) \, \gamma_{d-1} \, G_{-n } ( \omega_{p' },\vec {\bar  K}, \vec  
k)\Big ],
\end{split}
\end{equation}
where $\Omega_p $ and $ \omega_{p' } $ are bosonic and fermionic Matsubara 
frequencies respectively.
Now, it is easy to see that the leading contribution containing a pole in $1/\epsilon $ can be obtained from the terms where one frequency sum is replaced by an integral \cite{patel1,ips4} such that
\begin{align}
F_{fb} (T ) 
& =
 T
\Big [  
N_f   N_c \sum_{\omega_{p' }  } \int
 \frac{ dk_x \, dk_y\,  d^{ 1-\epsilon} \vec{ \bar K  }  }   {  (2 \, \pi)^{ 3-   \epsilon } } 
 \sum_{n=\pm } 
 \mathrm{Tr} \Big  [ G_n(\omega_{p' }, \vec{\bar K} , \vec k )   \,   \Sigma_{1,-n } ( \omega_{p' } , \vec{\bar K} , \vec k)
 \Big ] \nn
& \qquad  \quad    +  
g^2 \, (N_c^2 - 1)\, \mu^\epsilon \, \sum_{\Omega_p  } \int
 \frac{ dq_x \, dq_y\,  d^{ 1-\epsilon} \vec{ \bar Q} }   {  (2 \, \pi)^{ 3-   \epsilon } }
   \frac{  \Upsilon( \Omega_p , \vec{\bar Q }) }
   { \Omega_p^2 +  \vec {\bar Q}^ 2  + q_x^2+c^2 \, q_y^2  }
\Big ] .
\label{intenn}
\end{align}
Here, from Ref.~\cite{sur16},
\begin{align}
\Upsilon( \Omega_p , \vec{\bar Q } )
=\sum_{n=\pm } \int dk \, 
\mathrm{Tr}\Big  [\gamma_{d-1} \, G_{n} (k+q) \ \gamma_{d-1} \, G_{-n } (k)\Big ]
\vert_{ \text{singular}}
=-\frac{\left ( \Omega_p^2 + \vec{\bar Q }^2 \right )  ^ {\frac{2-\epsilon} {2} }}   
{16 \, \pi \,  v\, \epsilon} .
\end{align}

The first term in the square bracket gives the interaction correction to the fermionic part of the free energy.
Focussing on the fixed point, 
we set \(c=0\) in Eq.~(\ref{selfenf}) to obtain:
\begin{align}
\Sigma_{1,n } (k)&= 
\frac{i \,  g^2 \, \mu^\epsilon \, (N_c^2-1)  \left( \mathbf{\Gamma}\cdot\mathbf{K} \right)  }   
{8\, \pi   \, N_f \, N_c  \, \vert\mathbf{K}\vert^\epsilon   \, \epsilon}, \nonumber 
\end{align}
such that
\begin{align}
& F_{fb}^{( 1 )} (T) \nn
&= 
\frac{   g^2   \mu^\epsilon \, (N_c^2-1) \,T    }   
{ 8 \, \pi    \,v\,  \epsilon}
\sum_{\omega_p'  } \int
 \frac{ d\varepsilon_+(k) \,  d\varepsilon_-(k)\,  d^{ 1-\epsilon} \vec{ \bar K  }  }   {  (2 \, \pi)^{ 3-   \epsilon } } \,
 \frac{\mathrm{Tr}
\Big [  \lbrace \mathbf{\Gamma}\cdot\mathbf{K}+\gamma_{d-1}\ \varepsilon_- (k)\rbrace \,
(\mathbf{\Gamma}\cdot\mathbf{K}
) \Big ]}
{\big \lbrace \mathbf{K}^2+\varepsilon_+ ^2 (k)\big \rbrace  \, |\mathbf{K} |^{\epsilon}}\nn
&=\frac{   g^2  \mu^\epsilon \, (N_c^2-1)   \, \Gamma\left(1+\frac{\epsilon}{2}\right) }   
{ 4 \, \pi^2   \,v\,  \epsilon \, \Gamma\left(\frac{\epsilon}{2}\right)}
T \sum_{\omega_{p' } } 
\int
 \frac{ d\varepsilon_+(k) \,  d\varepsilon_-(k)\,  d^{ 1-\epsilon} \vec{ \bar K  }  }   {  (2 \, \pi)^{ 3-   \epsilon } }
  \int_0^1 dy\,y^{\frac{\epsilon} {2} -1}
  \, \frac{\mathbf{\bar{K}}^2+\omega_{p'}^2}
  {\Big [ \mathbf{\bar{K}}^2+ \omega_{p' } ^2+(1-y) \,\varepsilon_+^2 (k) \Big ]
  ^{1+ \frac{\epsilon} {2} }}\nn
&= \frac{   g^2   \mu^\epsilon \, (N_c^2-1) \,T  \,   \Gamma \left( \frac{\epsilon}{2}  \right )\,  \Gamma \left( \epsilon  \right ) \,  \Gamma \left( \frac{ 1- \epsilon}{2}  \right )\, \cos \left( \frac{\pi \, \epsilon}{2}  \right ) }   
{  2^{7-\epsilon } \,\pi^{ \frac{ 7 - \epsilon}{2}}  \, v}
 \sum_{\omega_{p' } } \frac{1}{\omega_{p'}^{2\,\epsilon - 2}   }
\int d\varepsilon_- (k)  \nn
&=\frac{   g^2   \mu^\epsilon \, (N_c^2-1) \,T^{3-2\,\epsilon }
\,   \Gamma \left( \frac{\epsilon}{2}  \right )\,  \Gamma \left( \epsilon  \right ) \,  \Gamma \left( \frac{ 1- \epsilon}{2}  \right )\, \cos \left( \frac{\pi \, \epsilon}{2}  \right )\, \zeta ( 2\,\epsilon - 2 , \frac{1}{2})  }   
{ 2^{ 4+\epsilon } \, \pi^{  \frac{3 \,(1+\epsilon)}{2} }  \,v }
\int d\varepsilon_- (k) \nn
&
=  \frac{  3\, g^2   \mu^\epsilon \, (N_c^2-1) \,T^{3-2\,\epsilon }\, \zeta (3)  }   
{  64 \, \pi^3   \,v\, \epsilon  }
\int d\varepsilon_- (k) \,.
\end{align}

Let us now compute the second term in the square bracket of Eq.~(\ref{intenn}), which gives the interaction correction to the bosonic part of the free energy as follows:
\begin{align}
F_{fb}^{(2)} (T)
&= -
\frac{ g^2 \, \mu^\epsilon  \, (N_c^2 - 1)\,T }
{ 16 \, \pi \,  v\, c\, \epsilon}
\sum_{\Omega_p  } \int
 \frac{ dq_x \, dq_y\,  d^{ 1-\epsilon} \vec{ \bar Q} }   {  (2 \, \pi)^{ 3-   \epsilon } }
  \frac{ \left ( \Omega_p^2 + \vec{\bar Q }^2 \right )  ^ {\frac{2-\epsilon} {2} }
  }   
{\Omega_p^2 +  \vec {\bar Q}^ 2  + q_x^2+  q_y^2 }
  \nn
&=
\frac{ g^2 \, \mu^\epsilon  \, (N_c^2 - 1)\,T }
{ 16 \, \pi \,  v\, c\, \epsilon}
\frac{ \cos \left( \frac{\pi \, \epsilon}{2}  \right )
\, \sec \left(  \pi \, \epsilon \right)
\,   \Gamma \left(2- \frac{\epsilon}{2}  \right )
\,  \Gamma \left(\epsilon- \frac{ 3 }{2}  \right ) }
{2^{7-\epsilon }  \, \pi^{  \frac{5-\epsilon}{2}   }
}
\sum_{\Omega_p  } \frac{1}{ \Omega_p^{2\, \epsilon -3} }\nn
&=
\frac{g^2 \, \mu^\epsilon\,  (N_c^2 - 1)\, \pi \, T^{4-2\,\epsilon}}
{720\, v \, c\, \epsilon}.
\end{align}

Finally, setting $\mu=1$, the free energy for the fermions at the fixed point, including the lowest order interaction, takes the form:
\begin{align}
F_f(T)& = F_f ^0(T) + F_{fb}^{(1)} (T) +\text{contribution from appropriate counterterms} \nn
&=
\frac{ 3\,N_f N_c\, T^{3-\epsilon} \,\zeta(3)}
{8\, \pi^2 \, v_*}  \int  d  \varepsilon_-(k)
+ \frac{  3\, g_*^2   \mu^\epsilon \, (N_c^2-1) \,T^{3-\epsilon }\,\ln T\, \zeta (3)  }   
{  64 \, \pi^3 \, v }
\int d\varepsilon_- (k) \nn
&=
\frac{ 3\,N_f N_c\, T^{3-\epsilon} \,\zeta(3)}
{8\, \pi^2 \, v_* }  \int  d  \varepsilon_-(k)
+ \frac{  3\,N_f N_c \, (z_\tau-1) \,T^{3-\epsilon }\,\ln T\, \zeta (3)  }   
{  8 \, \pi^2   \,v }
\int d\varepsilon_- (k) \nn
&=
\frac{ 3\,N_f N_c\, T^{4-\epsilon} \,\zeta(3)}
{4\, \pi^2 \, v_* }  \Big [ 1
+ T^{ \frac{1}{z_\tau}- \frac{1- z_ x }{z_\tau} - 1 + (z_\tau -1 )}\
\Big ] \,, 
\end{align}
near the hot-spots. Here we have used the fact that the temperature dependence for low $T$ can be estimated by putting $\pm T^{\frac{1}{z_\tau }}$ as the bounds of of the $\varepsilon_-(k)$ integral (which were
$\pm \omega  ^{\frac{1}{z_\tau} }  $ for the optical conductivity computation). This agrees with the expected scaling proportional to $ T^ {2 - \epsilon +\frac{1+z_x}{z_\tau}} $ to $\mathcal{O}(\epsilon)$, using the fact that $z_\tau = 1 + \mathcal{O} (\epsilon)$. Similarly, for the bosonic part, we get (using $g^2_*=\frac{8\, \pi N_c\, N_f}{(N_c^2-1)}\,(z_\tau-1)$):
\begin{align}
F_b(T)& = F_b^0(T) + F_{fb}^{(2)} (T) +\text{contribution from appropriate counterterms} \nn
&=
\frac{\pi^2  \, (N_c^2-1) \,T^{4-\epsilon}}
{ 90 \, c}
+  \frac{g_*^2 \,    (N_c^2 - 1)\, \pi \, T^{4- \epsilon}  \, \ln T }
{720\, v \, c} \nn
&=
\frac{\pi^2  \, (N_c^2-1)  \,T^{4-\epsilon}}
{ 90 \, c}
+  \frac{   \pi^2 \, (N_c^2 - 1) \, (z_\tau -1 )\,   T^{4- \epsilon-\frac{1-z_x}{ z_\tau }} \,  \ln T}
{ 45\,   c }  \nn
&=
\frac{\pi^2  \, (N_c^2-1)  \,T^{4-\epsilon}}
{ 90 \, c}  \Big [ 1
+ T^{   \frac{  z_ x -1  }{z_\tau}  + 2\, (z_\tau -1 )}
\Big ] \,,
\end{align}
near the hot-spots. Again, this agrees with the expected scaling proportional to $ T^ {2 - \epsilon +\frac{1+z_x}{z_\tau}} $ to $\mathcal{O}(\epsilon)$.

\section{Shear viscosity}
\label{etabys}

The momentum flux density tensor, or the stress tensor in short, is given by:
\begin{equation}
\label{eqn:texpr}
T_{\mu\nu}=\sum_M \Big [\frac{\delta \mathcal{L}}{\delta (\partial_\mu \zeta_M)}\partial_\nu \zeta_M
-\partial_{\mu}
\Big \lbrace  \frac{\delta  \mathcal L}  {\delta \left(  \partial_\alpha \partial^\alpha \zeta_M \right) }
\Big \rbrace  \partial_\nu \zeta_M
\Big ]
-\delta_{\mu \nu} \, \mathcal{L},
\end{equation}
where \(\zeta_M\) stands for all the quasiparticle fields in the theory and $\mathcal L$ is the Lagrangian density.
This gives us:
\begin{align}
T_{xy} (q) &=i \, v \sum_{j=1}^{N_f}  \sum_{\sigma=1}^{N_c} \sum_{ n=\pm}\int dk 
\, \bar{\Psi}_{n,j,\sigma }(k+q) \, \gamma_{d-1}\, \left(k_y + \frac{q_y}{2}\right)\Psi_{n,j,\sigma }(k),\\
T_{yx} (q)&=
i \sum_{j=1}^{N_f}  \sum_{\sigma=1}^{N_c} \sum_{ n=\pm} 
\int dk\, n\,  \bar{\Psi}_{n,j,\sigma }(k+q)\, \gamma_{d-1} \, \left(k_x + \frac{q_x}{2}\right)\Psi_{n,j,\sigma }(k).
\end{align}
We note that $ T_{xy} (q) \neq T_{yx} (q)$ due to the anisotropy of the model.

Using the Kubo formula \cite{Taylor}, the shear viscosity for flows along the $x$ and $y$ directions can be obtained from the formulae:
\be
 \eta_x (\omega)=\lim_{\mathbf{q}\to 0} \frac{1} {\omega}\,
 \, \chi_{T_{xy} \, T_{xy}}(\omega,\mathbf{q})\,, \quad 
 \eta_y (\omega)=\lim_{\mathbf{q}\to 0} \frac{1} {\omega}\,
 \, \chi_{T_{y x } \, T_{ y x }}(\omega,\mathbf{q})\,,
\ee
respectively, where
\be
 \chi_{T_{\alpha \beta } \, T_{ \alpha \beta}}(\omega,\mathbf{q})
 =  \langle T_{ \alpha \beta} \, T_{ \alpha \beta}\rangle (\omega, \vec q) 
\ee
is the autocorrelation function of the component $T_{\alpha \beta}$ of the stress tensor.

At one-loop order, we have
\begin{align}
& \langle T_{xy} \, T_{xy}\rangle (\omega) \nn
&=
 v^2\sum_{j=1}^{N_f}\sum_{\sigma=1}^{N_c}\sum_{n=\pm}
 \int dk\,k_y^2 \, \mathrm{Tr}\left[ \, \gamma_{d-1}\, G_n(k+q)\, \gamma_{d-1}\, G_n(k) \, \right] \nn
&=   \frac{4 \, v^2 N_f N_c  }  {(2 \, \pi)^{d+1} \, \Gamma \left( \frac{d-1} {2}\right)}
 \int_0^1 dt \int dk_x \, dk_y \,k_y^2
 \frac{  ( d- 2)\, \pi^{\frac{ d+1 } {2}}   }
 { \cos \left( \frac{ d \, \pi } {2}  \right) 
  \Big [ \varepsilon_+^2 (k) + t\, (1-t) \, \omega^2   \Big]^{\frac{3-d}{2}}    } \,, 
\end{align}
following the steps leading to Eq.~(\ref{jj1loop}). Here $q=(\omega, 0,\cdots, 0)$.
Changing the integration variables $(k_x, k_y) \rightarrow (\varepsilon_+ (k) , \varepsilon_-  (k) ) $,
we get:
\begin{align}
& \langle T_{xy} \, T_{xy}\rangle (\omega) \nn
&=   \frac{  v \, N_f N_c  }  {2 \,(2 \, \pi)^{d+1} \, \Gamma \left( \frac{d-1} {2}\right)}
 \int_0^1 dt \int d\varepsilon_+(k)\,d\varepsilon_-(k) \,\left( \varepsilon_+(k) -\varepsilon_-(k)  \right)^2
 \frac{  ( d- 2)\, \pi^{\frac{ d+1 } {2}}   }
 { \cos \left( \frac{ d \, \pi } {2}  \right) 
  \Big [ \varepsilon_+^2 (k) + t\, (1-t) \, \omega^2   \Big]^{\frac{3-d}{2}}    } \nn
&=
 \frac{   v  N_f N_c  }  {4\,(2 \, \pi)^{d+1}  }
 \int_0^1 dt \int  d\varepsilon_- (k)\, \varepsilon_-^2 (k)
 \frac{    \pi^{\frac{ d  } {2}} \, \Gamma \left ( 2 - \frac{d}{2}\right)  }
 { 
  \Big [   t\, (1-t) \, \omega^2   \Big]^{\frac{2-d}{2}}    }  \nn
 &= \frac{  (d-2)  \, v  N_f N_c \, \csc \left (  \frac{d \, \pi }{2}\right) \, \omega^{d-2} }
  { 2^{2 d+1}  \,  \pi^{\frac{ d-1  } {2}}  \, \Gamma \left ( \frac{d +1 }{2}\right)    }
\int d \varepsilon_- (k) \, \, \varepsilon_-^2 (k) \,.
\end{align}

The two-loop contribution to leading order in $c$ can also be found from the two-loop computation for $ \langle J_xJ_x\rangle (\omega )$ and using the fact that
\begin{align}
\int d \varepsilon_- (k) \, \, \varepsilon_-^2 (k)
=
\begin{cases}
\frac{  2 \,  \omega^{1+ \frac{ 3 }{z_\tau }-\epsilon } }  { 3    } 
+\mathcal{O}(\epsilon )   & \mbox{for regions close to hot-spots} \,, \\
\frac{ 2 \, k_F^3 \, \omega^{1 -\epsilon } }  { 3    }  + \mathcal{O}(\epsilon ) 
& \mbox{for regions far from hot-spots} \,.
\end{cases}
\end{align}

Finally, this leads to
\begin{align}
& \langle T_{xy} \, T_{xy}\rangle (\omega)
=\frac{   v_*\,  N_f N_c \,  \omega^{ 4-\epsilon } }  { 192 \, \pi    }
+ \frac{  v\,   N_f N_c \,
 (  z_\tau -1  ) \,  \omega^{1+\frac{3} {z_\tau }-\epsilon }}
{  192 \,  \pi   }
\ln \omega  \nn
&
=\frac{   v_* \,  N_f N_c \,  \omega^{ 4-\epsilon } }  { 192 \, \pi    }
\Big [ 1+
 (  z_\tau -1  ) \,  \omega^{ \frac{ 3 } {z_\tau }+\frac{ 1- z_x } { z_\tau}  -3  }
\ln \omega 
\Big ]  \nn
&
\simeq \frac{   v_* \,  N_f N_c \,  \omega^{ 4-\epsilon } }  { 192 \, \pi    }
\Big [ 1+
  \,  \omega^{ \frac{3\, (1- z_\tau )} {z_\tau }+\frac{ 1- z_x } { z_\tau}  +(  z_\tau -1  ) }
  \Big ]
\end{align}
near the hot-spots,
such that $\eta_x ( \omega ) \sim \omega^{  3-  \epsilon + \frac{ 1-z_x }{z_\tau }}  $ by the same argument as used for finding the scaling of optical conductivity.

Similarly, 
\begin{align}
&\langle T_{yx} \, T_{yx}\rangle (\omega )
=\frac{      N_f N_c \,  \omega^{ 4-\epsilon } }  { 192 \, \pi\, v_*^3    }
+ \frac{    N_f N_c \,
 (  z_\tau -1  ) \,  \omega^{1+\frac{3} {z_\tau }-\epsilon }}
{  192 \,  \pi \, v^3   }
\ln \omega  \nn
&
=\frac{     N_f N_c \,  \omega^{ 4-\epsilon } }  { 192 \, \pi \,  v_*^3   }
\Big [ 1+
 (  z_\tau -1  ) \,  \omega^{ \frac{ 3 } {z_\tau }-\frac{3\, ( 1- z_x ) } { z_\tau}  -3  }
\ln \omega 
\Big ]  \nn
&
\simeq \frac{   N_f N_c \,  \omega^{ 4-\epsilon } }  { 192 \, \pi  \,  v_*^3   }
\Big [ 1+
  \,  \omega^{ \frac{3\, (1- z_\tau )} {z_\tau } -\frac{ 3\, (1- z_x) } { z_\tau}  +(  z_\tau -1  ) }
  \Big ]
\end{align}
near the hot-spots,
such that $\eta_y ( \omega ) \sim \omega^{  3-  \epsilon - \frac{ 3 \, (1-z_x) }{z_\tau }}  $.

We will assume that the $T$-dependence of the DC viscosity
can be inferred from the frequency dependent shear viscosity by replacing $\omega $ by $T$ \cite{patel1,patel2}. 
Now, the entropy density ($s$), being the derivative of the free energy with respect to temperature $T$, scales as
\beq
s \sim T^{ 2-\epsilon +  \frac{z_x} {z_\tau } } \,,
\eeq
leading to
\beq
\eta_x/s \sim  T^{ 1 +  \frac{  1- 2 \,z_x} {z_\tau } } \,,
\quad
\eta_y/s \sim  T^{ 1 -  \frac{  3- 2 \,z_x} {z_\tau } } \,.
\eeq
Noting that $z_\tau = 1  +  \frac{ 2 \left ( N_c^2 -1 \right ) + N_c \, N_f   }
{2 \, \big \lbrace  
2 \left ( N_c^2 -3 \right ) + N_c \, N_f
\big \rbrace }
\, \epsilon $ and $z_x = 1 + \mathcal{O}  \left ( \epsilon^{4/3} \right )$ \cite{sur16}, we conclude that
\beq
\eta_x/s \sim  T^{- \frac{ 2 \left ( N_c^2 -1 \right ) + N_c \, N_f   }
{2 \, \big \lbrace  
2 \left ( N_c^2 -3 \right ) + N_c \, N_f
\big \rbrace }
\, \epsilon } \,,
\quad
\eta_y/s \sim  T^{\frac{3}{2} \frac{ 2 \left ( N_c^2 -1 \right ) + N_c \, N_f   }
{  2 \left ( N_c^2 -3 \right ) + N_c \, N_f}
\, \epsilon } \,.
\eeq
Hence, near the hot-spots, depending on whether $  \big \lbrace  2 \left ( N_c^2 -3 \right ) + N_c \, N_f \big \rbrace $ is positive or negative, the ratio $\eta_x/s$ or $\eta_y/s$
diverges at low $T$ with a negative power of $T$ proportional to $\epsilon$. Away from the hot-spots, we have $s \sim T^{2-\epsilon }$ and $\eta=\eta_x=\eta_y \sim T^{-\epsilon}$, so that $\eta /s \sim T^{-2} $ diverges at low temperatures.

\section{Superconducting fluctuations}
\label{pairing}

\begin{figure}
        \centering
        \subfloat[][]
                {\includegraphics[width=0.25 \textwidth]{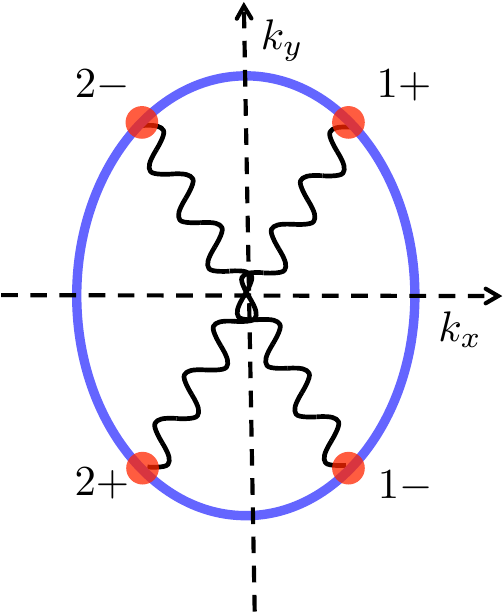} 
              { \label{cooper1}}} \qquad \qquad \qquad 
		\subfloat[][]
                {\includegraphics[width= 0.25 \textwidth]{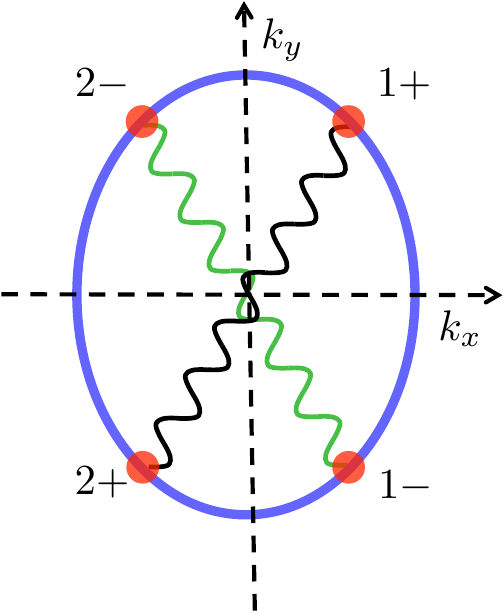}
		{\label{cooper2}}}
        \caption{\label{cooper}Cooper pairs with zero centre-of-mass momentum, such that the pairing gaps of the two hot-spot pairs have the same (opposite) sign(s), denoted by the same (different) colour(s) of the wiggly lines connecting them.}
\end{figure}

\begin{figure}
        \centering
        \subfloat[][]
                {\includegraphics[width=0.25 \textwidth]{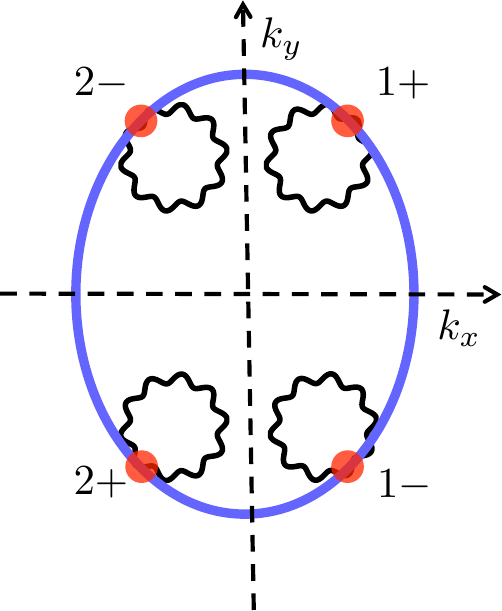} 
              { \label{fflo1}}} \qquad \qquad \qquad 
		\subfloat[][]
                {\includegraphics[width= 0.25 \textwidth]{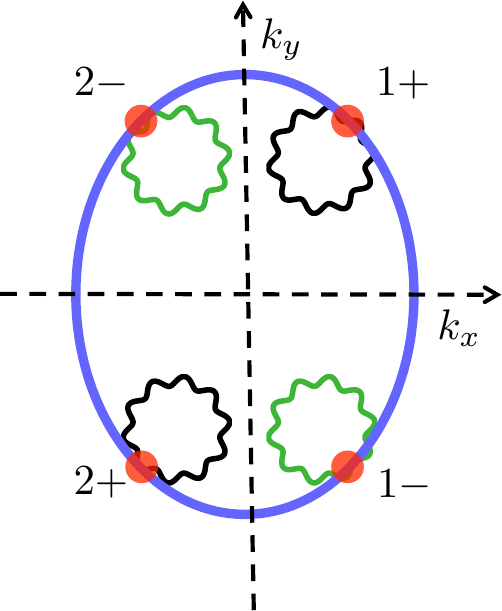}
		{\label{fflo2}}}
        \caption{\label{fflo}FFLO channels with $2\, k_F $ centre-of-mass momentum, such that the pairing gaps of the consecutive hot-spots have the same (opposite) sign(s), denoted by the same (different) colour(s) of the wiggly loops.}
\end{figure}

In this section, we will examine the potential superconducting instabilities both in the zero momentum and $ 2 \, k_F$ FFLO (Fulde-Ferrell-Larkin-Ovchinnikov) channels for the simplest case of spin-singlet pairing. Hence, we need $N_c\geq 2$. Hereafter, we will consider the case with $N_c= 2$. For Cooper pairs with zero centre-of-mass momentum, the pairing vertex is given by (Fig.~\ref{cooper}):
\begin{align}
S_1 = & \,  \mu \, V^0_+ \,  \sum_{j=1}^{N_f} \sum_{s,s'=1}^{2} \sum_{n = \pm} \int dk ~ 
\varepsilon_{s  s'}\, 
\Big [ \Psi_{n,j,s}^T ( - k) 
\, \gamma_{d-1}\,  \Psi_{n,j,s'}(k)
+ \bar   \Psi_{n,j,s} ( - k) 
\, \gamma_{d-1}\, \bar  \Psi_{n,j,s'}^ T (k)
\Big ] \nn
&+
\mu \, V^0_-\,  \sum_{j=1}^{N_f} \sum_{s,s'=1}^{ 2 } \sum_{n = \pm} \int dk ~ n \, \varepsilon_{s  s'}\, 
\Big [ \Psi_{n,j,s}^T ( - k) 
\, \gamma_{d-1}\,  \Psi_{n,j,s'}(k)
+ \bar   \Psi_{n,j,s} ( - k) 
\, \gamma_{d-1}\, \bar  \Psi_{n,j,s'}^ T (k)
\Big ] \,,
\end{align}
whereas for the FFLO case, with the centre of mass momentum equal to $2 \, k_F$, the pairing scenario can be captured by (Fig.~\ref{fflo}):
\begin{align}
S_2 = & \,  \mu \, V^{2k_F}_+ \, \sum_{j=1}^{N_f} \sum_{s,s'=1}^{2} \sum_{n = \pm} \int dk ~ 
\varepsilon_{s  s'}\, 
\Big [ \Psi_{n,j,s}^T ( - k) 
\,   \Psi_{n,j,s'}(k)
+ \bar   \Psi_{n,j,s} ( - k) 
\,   \bar  \Psi_{n,j,s'}^ T (k)
\Big ] \nn
&+
\mu \, V^{2k_F}_- \,  \sum_{j=1}^{N_f} \sum_{s,s'=1}^{ 2 } \sum_{n = \pm} \int dk ~ n \, \varepsilon_{s  s'}\, 
\Big [ \Psi_{n,j,s}^T ( - k) 
\,   \Psi_{n,j,s'}(k)
+ \bar   \Psi_{n,j,s} ( - k) 
\,  \bar  \Psi_{n,j,s'}^ T (k)
\Big ] \,.
\end{align}
The Cooper pairs formed from adjacent regions of the Fermi surface have the same or opposite signs depending on whether we are considering $ \lbrace V^0_+ ,\, V^{2k_F}_+ \rbrace $
or $\lbrace V^0_- , \,  V^{2k_F}_- \rbrace $.

The one-loop corrections for $S_1$ and $S_2$ are given by:
\bqa
\delta V^0_{n, \pm}  (k)&=& \pm
{ N_v\, \mu^{ 4- d} \,   V^0_{\pm} \,g^2   }
\int dq \, \gamma_{d-1}^T \, G^T_{-n } ( -k-q) \, \gamma_{d-1}
 \, {G_ {- n }}  (  k+q)\,   \gamma_{d-1} \, D(q) \,,\nn
 \delta V^{2 k_F}_{n, \pm}  (k)&=& \pm
\frac{ N_v\, \mu^{ 4- d} \,   V^0_{\pm} \,g^2   }
{N_f }
\int dq \, \gamma_{d-1}^T \, G^T_{-n } ( -k-q)
 \, {G_ {- n }}  (  k+q)\,   \gamma_{d-1} \, D(q) \,,
\eqa
respectively, where $N_v = \frac{2 \,(N_c +1)}{N_c N_f}$.

For $d = 3$, we have $\gamma_0^T = -\sigma_y =- \gamma_0$, $\gamma_1^T =  \sigma_z =  \gamma_ 1$ and $\gamma_2^T =  \sigma_x =  \gamma_ 1$. generalizing this to $ d= 3 -\epsilon$, we have the relations
\begin{align}
\gamma_0^T=- \gamma_0^T, \, \gamma_{\mu\neq 0}^T =  \gamma_{\mu\neq 0}.
\end{align}
In order to extract the divergent part, we can set $ \vec k=0  $, such that
\begin{align}
& \delta V^0_{n, \pm} (\vec K, \vec k =0 )  \nn
&= \mp \frac{ N_v\, \mu^{ 4- d} \,   V^0_{\pm} \,g^2  }
{i^2   }\nn
& \quad \times \, \int dq 
\frac{ \Big  [ (K_0 + Q_0)\,  \gamma_0 -\sum \limits _{\mu =1}^{d-2}(K_\mu + Q_\mu) \,
\gamma_\mu - \varepsilon_{ - n}( -q)\,  \gamma_{d-1}\Big  ]\, 
  \Big [  (\mbf{K} + 
\mbf{Q} ) \cdot \mbf{\Gamma} + \varepsilon_{- n}(  q)\,  \gamma_{d-1} 
\Big ]     \gamma_{d-1} }
{     \left ( \, \abs{\mbf{Q}}^2 + q_x^2 +  c^2 \, q_y^2 \,  \right ) ~ 
\lt[\,  \abs{\mbf{K} + \mbf{Q}}^2 +  \varepsilon_{- n}^2(  q) \, \rt]^2
} \nn
& = \pm N_v\, \mu^{ 4- d} \,   V^0_{\pm} \,g^2 
\int dq 
\frac{ \Big  [ (K_0 + Q_0) ^2  - \sum \limits _{\mu =1}^{d-2}(K_\mu + Q_\mu)^2   + \varepsilon_{ - n}^2(  q)\Big  ]  \gamma_{d-1} }
{ \left ( \abs{\mbf{Q}}^2 + q_x^2 +  c^2 \, q_y^2 \right ) ~ 
\lt[\,  \abs{\mbf{K} + \mbf{Q}}^2 +  \varepsilon_{- n}^2(  q) \, \rt]^2
}\nn
&\quad + \mbox{ terms not contributing to pairing}
\,.
\end{align}
We can look at the case of $c=0$ corresponding to the fixed point value and then the expression simplifies to
\begin{align}
& \delta V^0_{n, \pm} (\vec K, \vec k =0 )  
 \sim \pm N_v\, \mu^{ 4- d} \,   V^0_{\pm} \,g^2 
  \int dq 
\frac{ \Big  [ (K_0 + Q_0) ^2  - \sum \limits _{\mu =1}^{d-2}(K_\mu + Q_\mu)^2   + \varepsilon_{ - n}^2(  q)\Big  ]  \,   \gamma_{d-1} }
{ \left ( \abs{\mbf{Q}}^2 + \frac{( \varepsilon_ +   -  \varepsilon_ - )^2}  { 4 \, v^2 }   \right ) ~ 
\lt[\,  \abs{\mbf{K} + \mbf{Q}}^2 +  \varepsilon_{- n}^2(  q) \, \rt]^2
} 
\nn &
\sim  \pm \frac{  N_v\, \mu^{ 4- d} \,   V^0_{\pm} \,g^2 \,|\vec Q|^{\epsilon}   }
{16 \, \pi  \, \epsilon } \,.
\end{align}

Similarly, for the FFLO case, we have:
\begin{align}
& \delta V^{2 k_F}_{n, \pm} (\vec K, \vec k =0 )  \nn
&= \pm \frac{ N_v\, \mu^{ 4- d} \,   V^0_{\pm} \,g^2   }
{i^2   }
 \int dq
\frac{   \Big  [  (K_0 + Q_0)\,  \gamma_0 -\sum \limits _{\mu =1}^{d-2}(K_\mu + Q_\mu) \,
\gamma_\mu - \varepsilon_{ - n}( -q)\,  \gamma_{d-1}\Big  ]\, 
  \Big [  (\mbf{K} + 
\mbf{Q} ) \cdot \mbf{\Gamma} - \varepsilon_{- n}(  q)\,  \gamma_{d-1} 
\Big ]     \gamma_{d-1} }
{     \left ( \, \abs{\mbf{Q}}^2 + q_x^2 +  c^2 \, q_y^2 \,  \right ) ~ 
\lt[\,  \abs{\mbf{K} + \mbf{Q}}^2 +  \varepsilon_{- n}^2(  q) \, \rt]^2
} \nn
& = \mp N_v\, \mu^{ 4- d} \,   V^0_{\pm} \,g^2 
\int dq 
\frac{ \Big  [ (K_0 + Q_0) ^2  - \sum \limits _{\mu =1}^{d-2}(K_\mu + Q_\mu)^2   - \varepsilon_{ - n}^2(  q)\Big  ]  \gamma_{d-1} }
{ \left ( \abs{\mbf{Q}}^2 + q_x^2 +  c^2 \, q_y^2 \right ) ~ 
\lt[\,  \abs{\mbf{K} + \mbf{Q}}^2 +  \varepsilon_{- n}^2(  q) \, \rt]^2
}  + \mbox{ terms not contributing to pairing}
\,.
\end{align}
Again, simplifying to the case of $c=0$ corresponding to the fixed point value, we get:
\begin{align}
& \delta V^{2 k_F}_{n, \pm} (\vec K, \vec k =0 )  
 \sim \mp N_v\, \mu^{ 4- d} \,   V^0_{\pm} \,g^2 
  \int dq 
\frac{ \Big  [ (K_0 + Q_0) ^2  - \sum \limits _{\mu =1}^{d-2}(K_\mu + Q_\mu)^2   -\varepsilon_{ - n}^2(  q)\Big  ]  \,   \gamma_{d-1} }
{ \left ( \abs{\mbf{Q}}^2 + \frac{( \varepsilon_ +   -  \varepsilon_ - )^2}  { 4 \, v^2 }   \right ) ~ 
\lt[\,  \abs{\mbf{K} + \mbf{Q}}^2 +  \varepsilon_{- n}^2(  q) \, \rt]^2} 
\sim  \pm \frac{  N_v\, \mu^{ 4- d} \,   V^0_{\pm} \,g^2  \,|\vec Q|^{\epsilon}    }
{16 \, \pi  \, \epsilon } \,.
\end{align}

We can now write down the counterterms as:
\begin{align}
S_1^{\text{CT}  } = & -\frac{  \mu \, V^0_+ \, N_v  \,g^2} {16 \, \pi  \, \epsilon}  
\sum_{j=1}^{N_f} \sum_{s,s'=1}^{2} \sum_{n = \pm} \int dk ~ 
\varepsilon_{s  s'}\, 
\Big [ \Psi_{n,j,s}^T ( - k) 
\, \gamma_{d-1}\,  \Psi_{n,j,s'}(k)
+ \bar   \Psi_{n,j,s} ( - k) 
\, \gamma_{d-1}\, \bar  \Psi_{n,j,s'}^ T (k)
\Big ] \nn
&
-\frac{  \mu \, V^0_- \, N_v \,g^2} {16 \, \pi  \, \epsilon}  
   \sum_{j=1}^{N_f} \sum_{s,s'=1}^{ 2 } \sum_{n = \pm} \int dk ~ n \, \varepsilon_{s  s'}\, 
\Big [ \Psi_{n,j,s}^T ( - k) 
\, \gamma_{d-1}\,  \Psi_{n,j,s'}(k)
+ \bar   \Psi_{n,j,s} ( - k) 
\, \gamma_{d-1}\, \bar  \Psi_{n,j,s'}^ T (k)
\Big ] \,,
\end{align}
and
\begin{align}
S_2^{\text{CT}  } = & -\frac{  \mu  \, V^{2k_F}_+ \, N_v  \,g^2} {16 \, \pi  \, \epsilon}  
  \, \sum_{j=1}^{N_f} \sum_{s,s'=1}^{2} \sum_{n = \pm} \int dk ~ 
\varepsilon_{s  s'}\, 
\Big [ \Psi_{n,j,s}^T ( - k) 
\,   \Psi_{n,j,s'}(k)
+ \bar   \Psi_{n,j,s} ( - k) 
\,   \bar  \Psi_{n,j,s'}^ T (k)
\Big ] \nn
&
-\frac{  \mu  \, V^{2k_F}_- \, N_v  \,g^2} {16 \, \pi  \, \epsilon}   \sum_{j=1}^{N_f} \sum_{s,s'=1}^{ 2 } \sum_{n = \pm} \int dk ~ n \, \varepsilon_{s  s'}\, 
\Big [ \Psi_{n,j,s}^T ( - k) 
\,   \Psi_{n,j,s'}(k)
+ \bar   \Psi_{n,j,s} ( - k) 
\,  \bar  \Psi_{n,j,s'}^ T (k)
\Big ] \,.
\end{align}
The anomalous dimension for each of the $ \lbrace V^0_+ ,\, V^{2k_F}_+ ,\, V^0_- , \,  V^{2k_F}_- \rbrace $ is given by:
\begin{align}
\eta_v = z_\tau \left( \frac{g }{2} \,\partial_g  + u_{i ;0} \, \partial_{u_{i ;0}}
\right) \left( Z_{3,1} - Z_{v,1} \right) \,,
\end{align}
where $  Z_{v,1} =-  \frac{   N_v  \,g^2} {16 \, \pi  }   $ and $Z_{3,1} \propto c^2 \simeq 0$ is the counterterm coefficient associated with the $\gamma_{d-1}$ term of the one-loop fermion self-energy \cite{sur16}. This shows that $\eta_v >0$ and hence the coupling with the boson enhances the superconducting instability at the hot-spots. Furthermore, the FFLO pairing is found to be as strong as the zero-momentum pairing.

\section{Summary and outlook}
\label{summary}

For the $C_2$-symmetric SDW quantum critical point, our analysis has shown that the optical conductivity and free energy density obey the scaling relations expected from the anisotropy associated with the $x$ and $ y$ directions. Furthermore, this anisotropy leads to the observation that the direction-dependent $\eta/s$ ratios near the hot-spots are not universal numbers, unlike other strongly-coupled field theories.

We have also found that the fermion-boson coupling results in an enhancement of the instability of four-fermion interactions to superconducting pairing, both for the zero-momentum and $2 \, k_F$ Cooper pairs. However, such enhancement takes place only at the hot-spots and not on the entire Fermi surface. Hence, there will be no interpatch coupling term contributing to the beta functions of the pairing potentials, unlike the case of the Ising-nematic quantum critical point \cite{Max,ips-sc} (where it leads to a flow towards pairing instability irrespective of the initial value and sign of the four-fermion interaction strength).

A similar system to study in future works is the two-dimensional strange metal phase associated with the underlying quantum critical point in anisotropic electronic systems at the onset of inhomogeneous FFLO superconductivity \cite{phillip}.

\section{Acknowledgements}

We thank Denis Dalidovich, Andreas Eberlein, Aavishkar Patel and Shouvik Sur for helpful discussions. We are grateful to Andres Schlief for his valuable comments in improving the manuscript. This research was supported by NSERC of Canada, the Templeton Foundation and Perimeter Institute for Theoretical Physics. Research at Perimeter Institute is supported by the Government of Canada through the Department of Innovation, Science and Economic Development Canada and by the Province of Ontario through the Ministry of Research, Innovation and Science.

\appendix
\section{One-loop contribution to $\langle J_x J_x \rangle $}
\label{one-loop-jj}

The contribution to $\langle J_x J_x \rangle $ at one-loop level comes just from the free fermion part (see Fig.~\ref{one-loop-fig}) and is given by
\be
&& \langle J_x J_x \rangle_{\text{1-loop}} (\omega) = v^2  \sum_{j=1}^{N_f}  \sum_{\sigma=1}^{N_c} \sum_{ n=\pm} \int dk \,
\mathrm{Tr} \Big [ \gamma_{d-1} \, G_n (k+q) \,\gamma_{d-1} G_n( q) \Big] \nn
 &=& 2 \, v^2 N_f N_c  \int dk \,
\mathrm{Tr} \Big [ \gamma_{d-1} \, G_+ (k+q) \,\gamma_{d-1} G_+( q) \Big] \nn
 &=& 4 \, v^2 N_f N_c  \int dk \,
 \frac{ \varepsilon_+^2 (k) - \vec K \cdot \left(  \vec K + \vec Q  \right ) }
 { \Big [  \left(  \vec K + \vec Q  \right )^2  + \varepsilon_+^2 (k)   \Big] 
 \Big [     \vec K ^2  + \varepsilon_+^2 (k)   \Big]   } \nn
&=& 4 \, v^2 N_f N_c  \int_0^1 dt \int dk \,
 \frac{ \varepsilon_+^2 (k) - \vec K \cdot \left(  \vec K + \vec Q  \right ) }
 { \Big [  \left(  \vec K + t \,\vec Q  \right )^2  +t \,(1-t)   \,\vec Q^2+ \varepsilon_+^2 (k)   \Big]^2    } \nn
&=& -4 \, v^2 N_f N_c  \int_0^1 dt \int dk \,
 \frac{ \vec u^2 -t\, (1-t) \, \omega^2 -  \varepsilon_+^2 (k) }
 { \Big [  \vec u ^2  +t \,(1-t)   \,\omega^2+ \varepsilon_+^2 (k)   \Big]^2    } \nn
 &=&  \frac{4 \, v^2 N_f N_c  }  {(2 \, \pi)^{d+1} \, \Gamma \left( \frac{d-1} {2}\right)}
 \int_0^1 dt \int dk_x \, dk_y \,
 \frac{  ( d- 2)\, \pi^{\frac{ d+1 } {2}}   }
 { \cos \left( \frac{ d \, \pi } {2}  \right) 
  \Big [ \varepsilon_+^2 (k) + t\, (1-t) \, \omega^2   \Big]^{\frac{3-d}{2}}    } \,, 
  \label{jj1loop}
\ee
where $q=(\vec Q, \vec  q=0)$ and $\vec Q =(\omega, 0,\cdots, 0)$.
In the second last line, we have used Feynman parametrization and changed variables as $\vec u = \vec K + t \, \vec Q$. Finally, changing the integration variables $(k_x, k_y) \rightarrow (\varepsilon_+ (k) , \varepsilon_-  (k) ) $ and integrating over $ \varepsilon_+ (k)$, 
we get:
\be
\langle J_x J_x \rangle_{\text{1-loop}} (\omega) &=&
\frac{ 2 \, v  N_f N_c  }  {(2 \, \pi)^{d+1}  }
 \int_0^1 dt \int  d\varepsilon_- (k)
 \frac{    \pi^{\frac{ d  } {2}} \, \Gamma \left ( 2 - \frac{d}{2}\right)  }
 { 
  \Big [   t\, (1-t) \, \omega^2   \Big]^{\frac{2-d}{2}}    }  \nn
 &=& \frac{ 2 \, (d-2)  \, v  N_f N_c \, \csc \left (  \frac{d \, \pi }{2}\right) \, \omega^{d-2} }
  { 4^d  \,  \pi^{\frac{ d-1  } {2}}  \, \Gamma \left ( \frac{d +1 }{2}\right)    }
\int d \varepsilon_- (k)\,.
\ee
We note that the bounds on the integral over $\varepsilon_-(k) $ depends on whether we are considering the regions close to the hot-spots or far from the hot-spots. Hence 
\begin{align}
\label{hotspot0}
\int  d \varepsilon_+    (k)
=
\begin{cases}
2 \,\omega^{ \frac{1} {z_\tau } }    & \mbox{for regions close to hot-spots} \,, \\
 2 \,\lambda  
& \mbox{for regions far from hot-spots} \,,
\end{cases}
\end{align}
where $\lambda$ is a scale independent of $\omega$ and is of the order of $k_F \gg \omega $.

\section{Two-loop contributions to $\langle J_x J_x \rangle $}
\label{2loop-self}

In this appendix, we elaborate on the computation of the two-loop contributions to $\langle J_x J_x \rangle $.

\subsection{Contribution from the fermion self-energy correction}

We need to compute
\begin{align}
\langle J_x J_x\rangle_{\text{SE}}(q)&=2   \,  v^2\sum_{j=1}^{N_f}\sum_{\sigma=1}^{N_c}\sum_{n=\pm}\int dk\,\mathrm{Tr}\left[  \,   \gamma_{d-1}\, G_n(k)\, \Sigma_{1,  n} (k)\, G_n(k)\,
  \gamma_{d-1 }  \, G_n(k+ q) \, \right],\nonumber
\end{align}  
as shown in Eq.~(\ref{jjSE}).
The expression involves the two-loop contribution to $\langle J_x J_x \rangle $ coming from the fermion self-energy correction, which
is given by~\cite{sur16}
\begin{align}
\label{selfenf}
&\Sigma_{1,  n} (k)=
\frac{2 \, i \,\pi^{2-\frac{\epsilon} {2} }  \, \Gamma(\frac{\epsilon} {2} )}{(2 \, \pi)^{4-\epsilon}}
\frac{g^2\mu^\epsilon \left( N_c^2-1 \right )}{N_c\, N_f}
\int_0^1 dx 
\frac{\mathbf{\Gamma}\cdot\mathbf{K}
-\gamma_{d-1} \, \frac{c^2  \,  \varepsilon_{ - n}( k )}{c^2+x(1+v^2\, c^2-c^2)}}
{\left[\mathbf{K}^2+\frac{c^2  \, \varepsilon_{- n}^2(k )}{c^2+x\, (1+v^2\, c^2-c^2)}\right]^{\epsilon/2}}\frac{x^{-\frac{\epsilon} {2} }(1-x)^{\frac{1-\epsilon} {2} }}
{\sqrt{ c^2+x  \, (1+v^2\, c^2-c^2) } }.
\end{align}
Changing the integration variables as $(k_x, k_y) \rightarrow (\varepsilon_+ (k), \varepsilon_- (k))$, such that $d k_x \, dk_y = \frac{d \varepsilon_+ (k) \, d \varepsilon_-(k) } {2  \, v}$, we can immediately see that the term with the factor $\varepsilon_{- n}$ in the numerator of the integrand drops out on performing integrations leading to
$\langle J_x J_x \rangle_{\rm SE}$.
Hence the integral simplifies to
\begin{align}
& \langle J_x J_x \rangle_{\text{SE}}(\omega)\nn
&=  \frac{ 8\,v\,  ( N_c^2-1)\, \pi^{  2- \frac{ \epsilon} {2} }
\, \Gamma( \frac{\epsilon} {2} )\, g^2  \mu^\epsilon}{(2\pi)^{8-2\epsilon} }
\int_0^1 dx \frac{x^{-\frac{\epsilon} {2} }
(1-x)^{ \frac{1-\epsilon } {2} } } {\sqrt {c^2+x(1+v^2\, c^2-c^2) } } \nonumber \\
& \quad \times \int {d\varepsilon_+ (k)\, d\varepsilon_-(k) }\,  d^{2-\epsilon}\mathbf{K} 
\frac{
\mathbf{K}^4
+ \mathbf{K}^2 \left( \mathbf{Q}\cdot\mathbf{K} \right )
-\varepsilon_+^2 (k)
\, (3\, \mathbf{K}^2+\mathbf{K}\cdot\mathbf{Q})   }
{  \left [  \mathbf{K}^2+\varepsilon_+^2 (k) \right ]^2
\, \left [ \, (\mathbf{K}+\mathbf{Q})^2+\varepsilon_+^2(k)  \, \right  ]\,
\left[\mathbf{K}^2+\frac{c^2 \, \varepsilon_-^2 (k)  }{c^2+x\, (1+v^2\, c^2-c^2)}\right]^{\frac{\epsilon}{2}}
}\nn
&
=   \frac{ 16 \, v\,  ( N_c^2-1 )\, \pi^{2-   \frac{   \epsilon} {2} }
\, g^2  \mu^\epsilon \, \Gamma \left (\frac{ \epsilon  } {2} \right )}
{  (2\pi)^{8-2\epsilon}   }
\int_0^1 dx \, dy \, 
\frac {(1-x)^{  \frac{1-\epsilon } {2} }  \, (1-y)}  
{ x^{ \frac{\epsilon} {2} } \sqrt {c^2+x \, (1+v^2\, c^2-c^2) } }   \nonumber \\
& \quad \times \int   d\varepsilon_+(k) \, d\varepsilon_- (k) \,  d^{2-\epsilon}\mathbf{K} 
\, \frac{
\mathbf{K}^4
+ \mathbf{K}^2 \left( \mathbf{Q}\cdot\mathbf{K} \right )
-\varepsilon_+^2 (k)
\, (3\, \mathbf{K}^2+\mathbf{K}\cdot\mathbf{Q})
 }
{   
 \left [ \, (\mathbf{K}+y  \mathbf{Q})^2 + y \, ( 1-y ) \mathbf{Q}^2  +\varepsilon_+^2 (k) \, \right  ]^3 
 \,
\left[\mathbf{K}^2+\frac{c^2 \, \varepsilon_-^2 (k)}
{c^2+x\, (1+v^2\, c^2-c^2)}\right]^{\frac{\epsilon}{2}}  } \nn
&
= \frac{2\, v\,  ( N_c^2 -1 )\, \pi^{   3-\frac{ \epsilon} {2} }
\, g^2  \mu^\epsilon \, \Gamma \left (\frac{ \epsilon   } {2} \right )}
{  (2\pi)^{8-2\epsilon} }
\int_0^1 dx \, dy \, 
\frac {   (1-x)^{  \frac{1-\epsilon } {2} } \,  (1-y)  }  
{ x^{ \frac{\epsilon} {2} }  \sqrt {c^2+x(1+v^2\, c^2-c^2) }
}  \nn
&\quad\times\int \frac{ d \varepsilon_- (k) \,  d^{2-\epsilon}\mathbf{K}  }
 { 
\left[\mathbf{K}^2+\frac{c^2 \, \varepsilon_-^2 (k)}
{c^2+x\, (1+v^2\, c^2-c^2)}\right]^{\frac{\epsilon}{2}}
 } 
\Big [
\frac{ 3 \, \big \lbrace  \mathbf{K}^4
+ \mathbf{K}^2 \left( \mathbf{Q}\cdot\mathbf{K} \right ) \big \rbrace
 }
{   
 \big \lbrace \, (\mathbf{K}+y  \mathbf{Q})^2 + y \, ( 1-y ) \mathbf{Q}^2    \, \big \rbrace^{\frac { 5} { 2}  }   }
 -\frac{3\, \mathbf{K}^2+\mathbf{K}\cdot\mathbf{Q}
 }
 {   
 \big \lbrace \, (\mathbf{K}+y  \mathbf{Q})^2 + y \, ( 1-y ) \mathbf{Q}^2    \, \big \rbrace^{\frac { 3} { 2}  }   }
 \Big ], \label{eqn:SEint}
\end{align}
where $\vec Q =(\omega , 0, \cdots, 0) $.
Again, the bounds on the integral over $\varepsilon_-(k) $ depend on whether we are considering the regions close to the hot-spots or far from the hot-spots. Hence 
\begin{align}
\label{hotspot}
\int \frac{ d \varepsilon_-  (k)   }
 { 
\left[\mathbf{K}^2+\frac{c^2 \, \varepsilon_-^2 (k)}
{c^2+x\, (1+v^2\, c^2-c^2)}\right]^{\frac{\epsilon}{2}}  }
=
\begin{cases}
\frac{2 \,\omega^{ \frac{1} {z_\tau } }   }
{| \vec K | ^{\epsilon } } + \mathcal {O} (\epsilon)  & \mbox{for regions close to hot-spots} \,, \\
{2 \,\lambda^{ 1-\epsilon}   }
+ \mathcal {O} (\epsilon)
& \mbox{for regions far from hot-spots} \,.
\end{cases}
\end{align}

For the first case, we need to evaluate the integral
\begin{align}
I_1 (\omega) 
& \equiv 
\int_0^1 dx \, dy \, 
\frac {   (1-x)^{  \frac{1-\epsilon } {2} } \,  (1-y)  }  
{ x^{ \frac{\epsilon} {2} }  \sqrt {c^2+x(1+v^2\, c^2-c^2) }
}   \nn
&\,  \times  \int     \frac{d^{2-\epsilon}\mathbf{K} } { |\vec K|^{\epsilon } }
\Big [
\frac{ 3 \, \big \lbrace  \mathbf{K}^4
+ \mathbf{K}^2 \left( \mathbf{Q}\cdot\mathbf{K} \right ) \big \rbrace
 }
{   
 \big \lbrace \, (\mathbf{K}+y \, \mathbf{Q})^2 + y \, ( 1-y )\, \mathbf{Q}^2    \, \big \rbrace^{\frac { 5} { 2}  }   }
 -\frac{3\, \mathbf{K}^2+\mathbf{K}\cdot\mathbf{Q}
 }
 {   
 \big \lbrace \, (\mathbf{K}+y \, \mathbf{Q})^2 + y \, ( 1-y )\, \mathbf{Q}^2    \, \big \rbrace^{\frac { 3} { 2}  }   }
 \Big ].
\end{align}
We use another Feynman parameter in order to perform the integral over $\vec K$, such that
\begin{align}
I_1 (\omega) 
& = 
\int_0^1 dx \, dy \, dz\,
\frac {   (1-x)^{  \frac{1-\epsilon } {2} } \,  (1-y)  }  
{ x^{ \frac{\epsilon} {2} }  \sqrt {c^2+x \, (1+v^2\, c^2-c^2) }
} 
\frac{(   1-z)^{\frac{\epsilon}    {2}  -1}  }
{\Gamma  \left (  \frac{\epsilon}    {2} \right )  }   \nn
&\,  \times  
\int   d^{2-\epsilon}\mathbf{K} \,
 \Big [
\frac{ 3 \, \big \lbrace  \mathbf{K}^4
+ \mathbf{K}^2 \left( \mathbf{Q}\cdot\mathbf{K} \right ) \big \rbrace
 }
{   
 \big \lbrace \, (\mathbf{K}+y \, z \,  \mathbf{Q})^2 + y\, z\, ( 1-y\, z )\, \mathbf{Q}^2    \, \big \rbrace^{\frac { 5+\epsilon} { 2}  }   }
 -\frac{3\, \mathbf{K}^2+\mathbf{K}\cdot\mathbf{Q}
 }
 {   
 \big \lbrace \, (\mathbf{K}+y \, z\, \mathbf{Q})^2 +  y\, z\, ( 1-y\, z )\, \mathbf{Q}^2    \, \big \rbrace^{\frac { 3 +\epsilon} { 2}  }   }
 \Big ].
\end{align}
Shifting $\vec K \rightarrow \vec K - y \, z \, \vec Q$ and performing the integrals, we finally obtain:
\begin{align}
I_1 (\omega) 
&=-\frac{ 
\pi^2 \,   \omega^{1-2\epsilon}    }
{ 2 \left ( 1+ v^2\, c^2 -c^2 \right ) 
}\,
\Big[
(1+ v^2\, c^2 )\, \text{arccos} \left (\frac{c}  {  \sqrt { 1-v^2\, c^2  +v^2}   }   \right )
- c \,  \sqrt{1+ v^2\, c^2  }
\Big ] +\mathcal{O} \left ( \epsilon \right )\nn
& =- \frac{ 
\pi^3 \,   \omega^{1-2\epsilon}    }
{ 4} +\mathcal{O} \left ( c   \right ),
\end{align}
so that
\begin{align}
&
\langle J_x J_x \rangle_{\text{SE}}(\omega) =
-\frac{    v\,  ( N_c^2-1)\,
\, g^2 \, \omega^{1 +\frac{1} {z_\tau }-\epsilon } }
{  128\, \pi^2 \,   \epsilon  }
 \left (
\frac{\mu}
{\omega}
\right )^{\epsilon} ,
\end{align}
to leading order in $\epsilon $ and $c$, for the regions close to the hot-spots.

Now we consider the far from hot-spot case. Shifting \(\mathbf{K}\to\mathbf{K}-y\mathbf{Q}\) in (\ref{eqn:SEint}), we have:
\begin{align}
& \langle J_xJ_x\rangle_{\text{SE}}(\omega) \nn
&=\frac{4 \, v\, ( N_c^2 -  1)\,  \pi^{3-\frac{\epsilon}{2}}   \, g^2\mu^\epsilon \, \lambda^{1-\epsilon}
\,  \Gamma \left (\frac{\epsilon} {2 }  \right )}
{(2\pi)^{8-2\epsilon}}\int_0^1dx\,dy\,
\int d^{2-\epsilon}\mathbf{K} \,
\frac{(1-x)^{\frac{ 1-\epsilon }{2} }
\, (1-y)}{x^{ \frac{\epsilon} {2}}   \sqrt{c^2+x\, (1+v^2  \, c^2-c^2)}}
\nn
& 
\quad  \times 
\left[\frac{3   \, \left\{\left(\mathbf{K}-y\mathbf{Q}\right)^4+
 (  \, \mathbf{K}-y\mathbf{Q})^2
 \left (\mathbf{Q}\cdot(\mathbf{K}-y\mathbf{Q})   \right )\right\}}{\left\{\mathbf{K}^2+y\, (1-y)\mathbf{Q}^2\right\}^{ \frac{5 } {2}}}
 \right.
 \left.-\frac{3(\mathbf{K}-y\mathbf{Q})^2+(\mathbf{K}-y\mathbf{Q})\cdot\mathbf{Q}}{\left\{\mathbf{K}^2
 +y \, (1-y)\mathbf{Q}^2\right\}^{   \frac{3 } {2}  }}\right].
\end{align}
Defining $\bar {\vec K}  = ( K_1, K_2, \ldots, K_{d-1} ) $ such that $\vec K = (K_0, \bar {\vec K}  )$, we obtain
\begin{align}
& \langle J_xJ_x\rangle_{\text{SE}} (\omega) \nn
&=\frac{4 \, v\, ( N_c^2 -  1)\,  \pi^{3-\frac{\epsilon}{2}}   \, g^2\mu^\epsilon \, \lambda^{1-\epsilon}
 \,  \Gamma \left (\frac{\epsilon} {2 }  \right )}
 {(2\pi)^{8-2\epsilon}}
 \int_0^1 dx\,dy  \int d K_0 \,d^{ 1-\epsilon}\mathbf{\bar{K}} \, 
\frac{(1-x)^{\frac{ 1-\epsilon }{2} }
\, (1-y)}{x^{ \frac{\epsilon} {2}}   \sqrt{c^2+x\, (1+v^2  \, c^2-c^2)}}  \nn
&\quad \times 
\Big [
\frac{  
3 \, \big \lbrace  \bar{\vec {K}  }^2
+  \left ( K_0 -y \, \omega \right )^2 \big  \rbrace^2 +
\big \lbrace \mathbf{\bar{K}}^2 
+ \left ( K_0 -y \, \omega \right )^2 \big \rbrace 
(K_0 \, \omega-y\, \omega^2) 
}
{  \left\{\mathbf{\bar{K}}^2+K_0^2
+y \, (1-y) \,\omega^2\right\}^{5/2}    } \nn
& \quad \qquad 
-\frac{3 \, \big \lbrace \mathbf{\bar{K}}^2
+ \left ( K_0^2-y \, \omega \right )^2   \big \rbrace  +K_0 \, \omega-y \, \omega^2}{\left\{\mathbf{\bar{K}}^2+K_0^2+y \, (1-y) \,\omega^2\right\}^{\frac{3}{2}}}
\Big ] \nn
&=\frac{8 \, v\, ( N_c^2 -  1)\,  \pi^{3-\frac{\epsilon}{2}}   \, g^2\mu^\epsilon \, \lambda^{1-\epsilon}
 }
 {(2\pi)^{8-2\epsilon} \, \epsilon} \times 
\frac{2^{2-d} \, (d-2)  \, \pi^{\frac{1+d} {2}}  \, \omega^{d-2}}{\Gamma\left(\frac{1+d}{2}\right)\sin\left(\frac{d \, \pi}{2}\right)}
\int_0^1
 \frac{  dx\,  (1-x)^{\frac{ 1-\epsilon }{2} }
}
{x^{ \frac{\epsilon} {2}}   \sqrt{c^2+x\, ( 1+v^2  \, c^2-c^2  )}}   \nn
&= \frac{8 \, v\, ( N_c^2 -  1)\,  \pi^{3-\frac{\epsilon}{2}}   \, g^2\mu^\epsilon \, \lambda^{1-\epsilon}
 }
 {(2\pi)^{8-2\epsilon}  \, \epsilon} \times 
\frac{2^{2-d} \, (d-2)\,\pi^{\frac{1+d} {2} }\,\omega^{ 1- \epsilon }}
{\Gamma\left(\frac{1+d}{2}\right)\sin\left(\frac{d \, \pi}{2}\right)}
\nn &\quad\times
 \frac{ c^3 \left(v^2-1\right)+\left( 1  + v^2 \, c^2\right) \sqrt{ 1+v^2  \, c^2-c^2  }  \, 
\text{arctan}
\left(  \frac {\sqrt{  1+v^2  \, c^2-c^2   }   }   {c}  \right)
+ c  }
{\left(    1+v^2  \, c^2-c^2     \right  )^2}   \nn
&=-\frac{g^2\,  ( N_c^2  -1)   \, v \, \lambda \, \omega^{1-\epsilon}}
{128  \, \pi^2 \, \epsilon}  \left ( \frac{ \mu }  {\lambda }    \right ) ^\epsilon\,,
\end{align}
to leading order $c$ and $\epsilon $.


\subsection{Contribution from the vertex correction}
The contribution to $\langle J_x J_x\rangle  $ from the vertex correction can be written as:
\beq
\label{jjvert10}
\langle J_x J_x\rangle_{\text{VC}}(q)= i\,v ^2 \sum_{j=1}^{N_f}\sum_{\sigma=1}^{N_c}\sum_{n=\pm}  \int dk\,\mathrm{Tr}\left[ \, \gamma_{d-1}G_n(k )\, \Xi_{n }( k )\, G_n(k+q) \, \right], 
\eeq
where $ \Xi_{n }(k)$ is the one-loop fermion-boson vertex correction.
Now, from a Ward identity \cite{metlsach}, we have:
\[
\left.\Xi_{n }(k)\right\vert_{\text{singular}}=\left.-n\, \frac{\partial\Sigma_{1,n }(k)}
{\partial k_y }\right\vert_{\text{singular}},\]
Inserting this into Eq.~(\ref{jjvert10}), we get:
\begin{align}
&\langle J_x J_x\rangle_{\text{VC}}(\omega) \nn
&=  
\frac{2 \, \pi^{2-\frac{\epsilon}{2} }  \,  \Gamma\left(\frac{\epsilon}{2}\right) 
\,  v^2   g^2  \mu^\epsilon (N_c^2-1) \, c^2}
{(2\pi)^{4-\epsilon} }
\sum_{n=\pm}\int dk\int_0^1 dx\,\frac{
 (1-x)^{\frac{1-\epsilon} {2} }
 }
{\lbrace  c^2+x\, (1+v^2\, c^2-c^2)\rbrace  ^{\frac{3}{2}}
\,x^{\frac{ \epsilon} {2} }}
\nn & \qquad \quad  \times \, 
\frac{\mathrm{Tr}
\left[
\, \gamma_{d-1} \, \lbrace 
\, \mathbf{\Gamma}\cdot \vec Q  +\gamma_{d-1}\, \varepsilon_n (k)
\, \rbrace 
\, \gamma_{d-1} \, \lbrace \,  \mathbf{\Gamma}\cdot (\mathbf{K}+\mathbf{Q})
+\gamma_{d-1}  \, \varepsilon_n (k) \, \rbrace \, \right]
}
{\big [  \mathbf{K}^2+\frac{c^2\, \varepsilon^2 _{ -n }( k ) }
{c^2+x \, (1+v^2\, c^2-c^2)}   \big ]^{ \frac{\epsilon}{2}   } \,
\big [   \mathbf{K}^2+   \varepsilon^2 _{ n }( k )   \big ] \,
\big [  (\mathbf{K}+\mathbf{Q})^2+  \varepsilon^2 _{ n }( k )   \big ]    
} , \nonumber
\end{align}
where $\vec Q=(\omega , 0, \cdots, 0)$.
Using
\[\mathrm{Tr}
\left[
\, \gamma_{d-1} \, \lbrace 
\, \mathbf{\Gamma}\cdot(\mathbf{K}+\mathbf{Q})+\gamma_{d-1}\, \varepsilon_n (k)
\, \rbrace 
\, \gamma_{d-1} \, \lbrace \,  \mathbf{\Gamma}\cdot \mathbf{K}
+\gamma_{d-1}  \, \varepsilon_n (k) \, \rbrace \, \right]=
2\, \big [ \varepsilon_n^2 (k ) -\mathbf{K}\cdot(\mathbf{K}+\mathbf{Q})
\big ]  ,\]
we obtain:
\begin{align}
& \langle J_xJ_x\rangle_{\text{VC}}(\omega) \nn
&=
\frac{ 4 \, v  \, \pi^{2-\frac{\epsilon}{2}}
g^2 \, \mu^\epsilon \, (N_c^2-1)\,c^2\, \Gamma \left( \frac{\epsilon}{2} \right)  }
{ (2\pi)^{ 8 -2\, \epsilon}   }
\int d^{2-\epsilon}\mathbf{K} 
\int d\varepsilon_+(k ) \,d\varepsilon_-(k)  
\int_0^1 dx\,
\frac{
 (1-x)^{\frac{1-\epsilon} {2} }
 }
{\lbrace  c^2+x\, (1+v^2\, c^2-c^2)\rbrace  ^{\frac{3}{2}}
\,x^{\frac{ \epsilon} {2} }} \nn
& \qquad \qquad \qquad \qquad \qquad  \qquad  \times \,
\frac{\varepsilon_+^2 (k)  -\mathbf{K}\cdot(\mathbf{K}+\mathbf{Q})}
{\big [ \mathbf{K}^2+\varepsilon_+ ^ 2 (k) \big ] \,
\big [  (\mathbf{K}+\mathbf{Q})^2+\varepsilon_+^2 (k)  \big ] \, 
\big [ \mathbf{K}^2+\frac{c^2  \, \varepsilon_-^2 (k)}
{c^2+x \, (1+v^2\, c^2-c^2)}\big ]^{\frac{\epsilon} {2}  }} \,.\nonumber
\end{align}
Applying Feynman parametrization and
carrying out the integral over \(\varepsilon_+ (k) \) gives us:
\begin{align}
& \langle J_xJ_x\rangle_{\text{VC}}(\omega)\nn
&=
 \frac{ 2 \, v  \, \pi^{3-\frac{\epsilon}{2}}
g^2 \,  \mu^\epsilon  \,(N_c^2-1) \,c^2\,  \Gamma \left( \frac{\epsilon}{2} \right)}
{ (2\pi)^{ 8 -2\, \epsilon}}
\int d^{2-\epsilon}\mathbf{K}
\int d\varepsilon_-(k)  
\int_0^1 dx\, dy \,
\frac{
 (1-x)^{\frac{1-\epsilon} {2} }
 }
{\lbrace  c^2+x\, (1+v^2\, c^2-c^2)\rbrace  ^{\frac{3}{2}}
\,x^{\frac{ \epsilon} {2} }}
\nn
&\qquad  \qquad \qquad  \qquad \qquad \qquad
\times\frac{ y\, \mathbf{Q}^2  +   \mathbf{K}\cdot \mathbf{Q}  \, (2 \, y-1) }
{
\big [ \left ( \mathbf{K} + y \, \vec Q  \right )^2 + y \, (1-y) \mathbf{Q}^2
\big ]^{ \frac{3} {2} }   \,
\big [ \mathbf{K}^2+\frac{c^2  \, \varepsilon_-^2 (k)}
{c^2+x \, (1+v^2\, c^2-c^2)}\big ]^{\frac{\epsilon} {2}  }
}. 
\end{align}
Now we need to do the integral over \(\varepsilon_-(k) \) using Eq.~(\ref{hotspot}).

For regions near the hot-spots, we get
\begin{align}
& \langle J_xJ_x\rangle_{\text{VC}}(\omega)\nn
&=
  \frac{ 4 \, v  \, \pi^{3-\frac{\epsilon}{2}}
g^2  \, \mu^\epsilon   \, (N_c^2-1)  \,c^2 \, \omega^{\frac{1}{z_{\tau }}}  }
{ (2\pi)^{ 8 -2\, \epsilon}}
\int d^{2-\epsilon}\mathbf{K}\int_0^1
dx\, dy \,dz
\frac{
 (1-x)^{\frac{1-\epsilon} {2} }   \, z^{\frac{1}{2 } }  \, (1-z)^{ \frac{\epsilon}{2}-1}
 }
{\lbrace  c^2+x\, (1+v^2\, c^2-c^2)\rbrace  ^{\frac{3}{2}}
\,x^{\frac{ \epsilon} {2} }} \nn
&\qquad \qquad \qquad    \qquad \qquad \qquad \qquad  \times
\frac{\Gamma \left ( \frac{3 + \epsilon } {2}   \right )  }
{\Gamma\left ( \frac{3 } {2}   \right )   }
\,\frac{  y  \,  \mathbf{Q}^2+\mathbf{K}\cdot \mathbf{Q} \,  (2 \,y-1)   }
{
\big [  (\mathbf{K}+yz \,\mathbf{Q})^2+yz \, (1-yz)\mathbf{Q}^2
\big ]  ^{\frac{ 3+ \epsilon} {2 }}
}\nn
& = \frac{   v\,   g^2  \, (N_c^2-1) \, c \, \omega^{1+ \frac{1}{z_\tau}  - \epsilon}   }
{ 32 \,   \pi ^{ 3 } \, \epsilon} 
\left(   \frac{ \mu}{\omega}\right)^\epsilon+ \mathcal{O} (\epsilon ^0 ) \,.
\end{align}

For regions far from the hot-spots, we have
\begin{align}
\langle J_xJ_x\rangle_{\text{VC}}(\omega)&=
\frac{ 4 \, v \, \pi^{ 3-\frac{\epsilon}{2} }
\, \Gamma\left(\frac{\epsilon}{2}\right) \, g^2 \, \mu^\epsilon  \, (N_c^2-1)\, \lambda^{1-\epsilon}\, c^2}
{(2\pi)^{8-2\, \epsilon}}
\int d^{2-\epsilon}\mathbf{K}\int_0^1
dx\, dy \, \frac{
 (1-x)^{\frac{1-\epsilon} {2} }
 }
{\lbrace  c^2+x\, (1+v^2\, c^2-c^2)\rbrace  ^{\frac{3}{2}}
\,x^{\frac{ \epsilon} {2} }} \nn
&
\qquad \qquad  \qquad \qquad \qquad \qquad \qquad \qquad \qquad \times
\frac{ y \, \mathbf{Q}^2+\mathbf{K}\cdot\mathbf{Q}\, (2 \, y-1) }
{\big [ \left ( \mathbf{K} + y \, \vec Q  \right )^2 + y \, (1-y) \mathbf{Q}^2
\big ]^{ \frac{3} {2} }
}
\nn
&=
 \frac{   v
\,  g^2 \,   (N_c^2-1) \,c \,\lambda \, \omega^{1 - \epsilon } }
{  32 \, \pi^{ 3 } \, \epsilon } 
\left( \frac{\mu} {\lambda } \right)^\epsilon   + \mathcal{O} (\epsilon ^0 ) .
\end{align}

\bibliography{notes-c2}
\end{document}